\documentclass[manuscript,screen,acmtog]{acmart}

\AtBeginDocument{%
  }

\setcopyright{acmlicensed}
 \copyrightyear{2025} 
\acmYear{2025} 
\setcopyright{cc}
 \setcctype{by}
 \acmConference[SA Conference Papers '25]{SIGGRAPH Asia 2025 Conference
 Papers}{December 15--18, 2025}{Hong Kong, Hong Kong}
 \acmBooktitle{SIGGRAPH Asia 2025 Conference Papers (SA Conference Papers
'25), December 15--18, 2025, Hong Kong, Hong
 Kong}\acmDOI{10.1145/3757377.3763852}
 \acmISBN{979-8-4007-2137-3/2025/12}

\usepackage{algorithm}
\usepackage{algorithmic}
\usepackage{subfigure} 
\usepackage{framed}
\usepackage{xcolor}

\definecolor{myblue}{RGB}{230, 240, 255} % Light blue background
\definecolor{myborder}{RGB}{50, 100, 200} % Darker blue border

\definecolor{shadecolor}{gray}{0.9} % background of main box
\definecolor{titlebg}{gray}{0.8}    % title background
\definecolor{rulecolor}{gray}{0.7}  % horizontal rule

\begin{document}

%%
%% The "title" command has an optional parameter,
%% allowing the author to define a "short title" to be used in page headers.
\title{Off-Centered WoS-Type Solvers with Statistical Weighting}

%%
%% The "author" command and its associated commands are used to define
%% the authors and their affiliations.
%% Of note is the shared affiliation of the first two authors, and the
%% "authornote" and "authornotemark" commands
%% used to denote shared contribution to the research.
% \author{Ben Trovato}
% \authornote{Both authors contributed equally to this research.}
% \email{trovato@corporation.com}
% \orcid{1234-5678-9012}
% \author{G.K.M. Tobin}
% \authornotemark[1]
% \email{webmaster@marysville-ohio.com}
% \affiliation{%
%   \institution{Institute for Clarity in Documentation}
%   \city{Dublin}
%   \state{Ohio}
%   \country{USA}
% }

% \author{Lars Th{\o}rv{\"a}ld}
% \affiliation{%
%   \institution{The Th{\o}rv{\"a}ld Group}
%   \city{Hekla}
%   \country{Iceland}}
% \email{larst@affiliation.org}

\author{Anchang Bao}
\affiliation{%
  \institution{Tsinghua University}
  \city{Beijing}
  \country{China}
}
\email{baoanchang02@gmail.com}

\author{Jie Xu}
\authornote{Finished when visiting Tsinghua University.}
\affiliation{%
  \institution{University of Electronic Science and Technology of China}
  \city{Chengdu}
  \country{China}
  }
\email{xujie6338@gmail.com}

\author{Enya Shen}
\authornote{Corresponding author: Enya Shen}
\affiliation{%
  \institution{Tsinghua University}
  \city{Beijing}
  \country{China}
  }
\affiliation{%
  \institution{Haihe Lab of ITAI}
  \city{Tianjin}
  \country{China}
  }

\email{shenenya@tsinghua.edu.cn}

\author{Jianmin Wang}
\affiliation{%
  \institution{Tsinghua University}
  \city{Beijing}
  \country{China}
  }
\email{jimwang@tsinghua.edu.cn}
% \author{Aparna Patel}
% \affiliation{%
%  \institution{Rajiv Gandhi University}
%  \city{Doimukh}
%  \state{Arunachal Pradesh}
%  \country{India}}

% \author{Huifen Chan}
% \affiliation{%
%   \institution{Tsinghua University}
%   \city{Haidian Qu}
%   \state{Beijing Shi}
%   \country{China}}

% \author{Charles Palmer}
% \affiliation{%
%   \institution{Palmer Research Laboratories}
%   \city{San Antonio}
%   \state{Texas}
%   \country{USA}}
% \email{cpalmer@prl.com}

% \author{John Smith}
% \affiliation{%
%   \institution{The Th{\o}rv{\"a}ld Group}
%   \city{Hekla}
%   \country{Iceland}}
% \email{jsmith@affiliation.org}

% \author{Julius P. Kumquat}
% \affiliation{%
%   \institution{The Kumquat Consortium}
%   \city{New York}
%   \country{USA}}
% \email{jpkumquat@consortium.net}

%%
%% By default, the full list of authors will be used in the page
%% headers. Often, this list is too long, and will overlap
%% other information printed in the page headers. This command allows
%% the author to define a more concise list
%% of authors' names for this purpose.
\renewcommand{\shortauthors}{Bao et al.}
\newcommand{\changed}[1]{{#1}}

\newcommand{\changedd}[1]{{#1}}

%%
%% The abstract is a short summary of the work to be presented in the
%% article.
\begin{abstract}
Stochastic \changed{PDE} solvers have emerged as a powerful alternative to traditional discretization-based methods for solving partial differential equations (PDEs), especially in geometry processing and graphics. While off-centered estimators enhance sample reuse in \changed{WoS-type Monte Carlo solvers}, they introduce correlation artifacts and bias when Green’s functions are approximated. In this paper, we propose a statistically weighted off-centered \changed{WoS-type estimator} that leverages local similarity filtering to selectively combine samples across neighboring evaluation points. Our method balances bias and variance through a principled weighting strategy that suppresses unreliable estimators. We demonstrate our approach's effectiveness on various PDEs—including screened Poisson equations—and boundary conditions, achieving consistent improvements over existing solvers such as vanilla Walk on Spheres, mean value caching, and boundary value caching. Our method also naturally extends to gradient field estimation and mixed boundary problems.
\end{abstract}

%%
%% The code below is generated by the tool at http://dl.acm.org/ccs.cfm.
%% Please copy and paste the code instead of the example below.
%%


% \ccsdesc[500]{Do Not Use This Code~Generate the Correct Terms for Your Paper}
% \ccsdesc[300]{Do Not Use This Code~Generate the Correct Terms for Your Paper}
% \ccsdesc{Do Not Use This Code~Generate the Correct Terms for Your Paper}
% \ccsdesc[100]{Do Not Use This Code~Generate the Correct Terms for Your Paper}

%%
%% Keywords. The author(s) should pick words that accurately describe
%% the work being presented. Separate the keywords with commas.

% \received{20 February 2007}
% \received[revised]{12 March 2009}
% \received[accepted]{5 June 2009}

\begin{teaserfigure}
    \centering
    \includegraphics[width=0.95\textwidth]{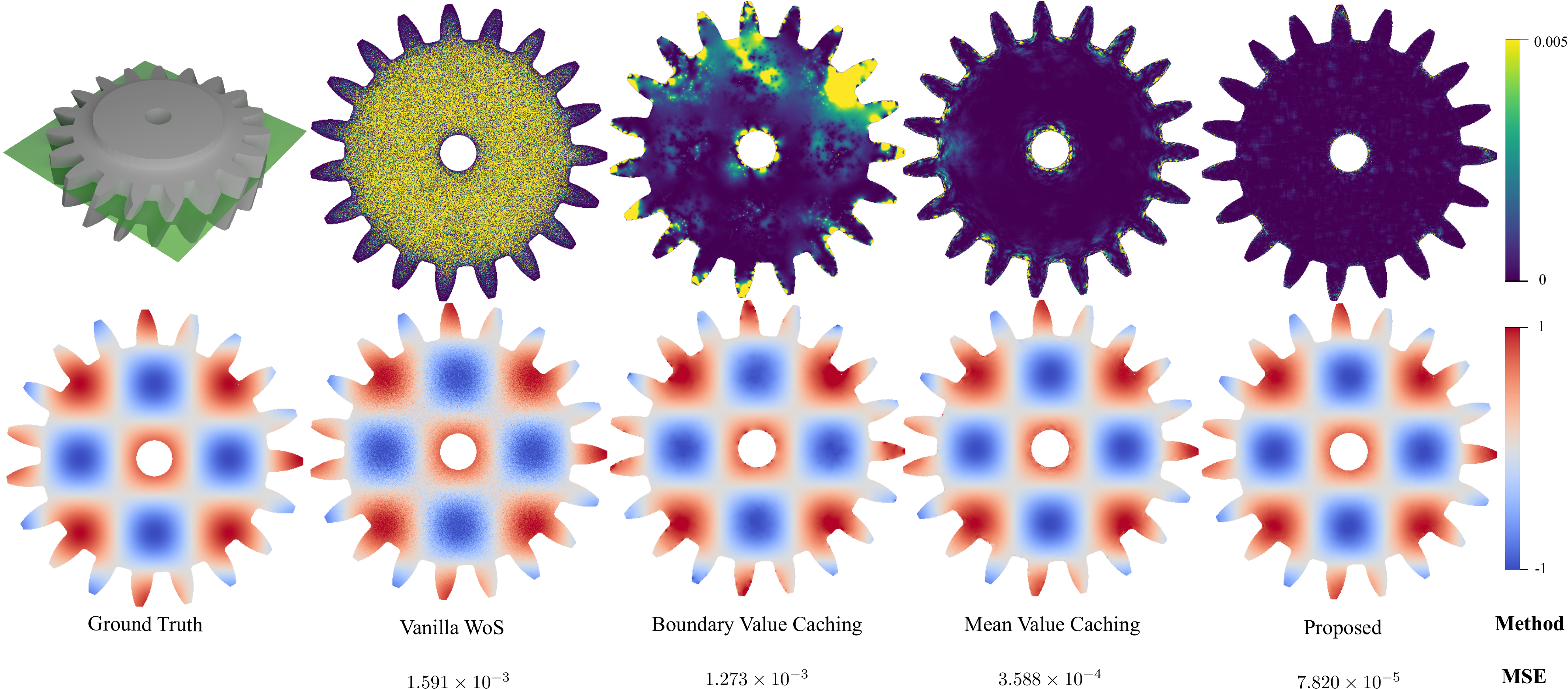}
    \caption{We propose a statistical weighting strategy for off-centered Monte Carlo estimators, demonstrating remarkable variance reduction results. The first row demonstrates error maps, and the results are shown in the second row. Compared with both vanilla walk on sphere algorithm and different variance reduction methods, our method reaches better mean square error within the same time.}
    \label{fig:teaserfigure}
\end{teaserfigure}

%%
%% This command processes the author and affiliation and title
%% information and builds the first part of the formatted document.
\maketitle

\section{Introduction}

In recent years, Monte Carlo partial differential equation (PDE) solvers have significantly influenced the field of computer graphics. These solvers have been applied in various areas, including geometry processing \cite{sawhney2020monte}, fluid simulation \cite{rioux2022monte, sugimoto2024velocity}, and cage-based deformation \cite{de2024stochastic}. Unlike traditional PDE solvers based on the Finite Element Method (FEM), \changed{stochastic PDE solvers} eliminate the need for spatial discretization and offer an efficient methodology for solving boundary value problems. Several \changed{stochastic PDE solvers} have been developed to address different types of PDEs, such as the Laplace, Poisson, and screened Poisson equations, under various boundary conditions, including Dirichlet \cite{sawhney2020monte}, Neumann \cite{sawhney2023walk}, and Robin \cite{miller2024walkin}. \citet{sawhney2022grid} further extended these solvers to handle spatially varying coefficients. More recently, differentiable stochastic PDE solvers have been introduced to address inverse problems \cite{miller2024differential, yu2024differential, yilmazer2024solving}.

In the \changed{WoS-type} Monte Carlo solver proposed by \citet{sawhney2020monte}, estimations at each evaluation point are independent. Variance reduction methods such as control variates and importance sampling are employed to accelerate the convergence of pointwise estimators. To further enhance convergence by leveraging the underlying PDE structure, recent works \cite{qi2022bidirectional, miller2023boundary, bakbouk2023mean, czekanski2024walking} have introduced correlations between evaluation points using various strategies, achieving remarkable variance reduction. 

An interesting acceleration approach is the off-centered estimator, which has been studied in some cases. \citet{hwang2015off, yu2024differential} use the off-centered estimator to accelerate the random walk from evaluation points near the domain boundary. For Laplace equations, \citet{czekanski2024walking} used the off-centered estimator to introduce correlation between evaluation points and \changed{used a deterministic bound} to weight the off-centered estimators.

However, when adapting the approach of \citet{czekanski2024walking} to a more general setup, we found that the simple off-centered estimator has the following two issues:
\begin{itemize}
    \item \textit{Correlation artifacts}. If one reused sample has a \changed{relatively large error}, it can affect all evaluation points that depend on it, introducing large \changed{errors} and artifacts.
    \item \textit{Bias}: If the off-centered Green's function cannot be computed efficiently and must be approximated, additional bias may be introduced.
\end{itemize}
To mitigate these issues, we adopt a statistical weighting scheme inspired by recent Monte Carlo denoising techniques \cite{sakai2024statistical}, allowing us to suppress extreme points and providing a bias-variance trade-off when using approximated Green's functions.

Our main contributions are as follows.
\begin{itemize}
    \item An efficient off-centered \changed{WoS-type estimator} that aggregates information across multiple integration domains using a principled, statistically-weighted combination.
    \item Extension of the off-centered estimator to a wide class of PDE problems, including those with screened Poisson equations, mixed Dirichlet–Neumann boundary conditions, and gradient field estimation.
    \item A comprehensive experimental evaluation showing that our method consistently achieves lower variance and error compared to state-of-the-art baselines.
\end{itemize}

\section{Related Works}

\subsection{Stochastic PDE Solver}

The theoretical foundation of stochastic PDE solvers stems from potential theory and probability. \changed{Stochastic PDE solvers can be applied to a wide range of scenarios; for clarity of exposition, we briefly introduce the simplest Laplace equation case.} \citet{kakutani1944143} established the key result: 
\begin{equation}
    u(x) = \mathbb{E}(g(X_\tau))
\end{equation}
where $X$ is the Brownian motion starting from $X(0) = x$ and $\tau = \inf\{t : X(t) \in \partial \Omega\}$ is the first-hit time to boundary. 

This formulation offers a stochastic approach to solving linear elliptic PDEs: at each point inside the domain, Brownian random walks are simulated, and the solution is estimated by averaging the boundary values at the hitting points. However, explicit simulation is computationally expensive due to the need for small time steps and frequent boundary checks in the temporal difference.

To address this bottleneck, \citet{muller1956some} proposed the Walk on Spheres (WoS) algorithm, which implicitly simulates Brownian motion. At each step of the WoS algorithm, the distance $d$ between the current position $p$ and the \changed{given boundary $\partial \Omega$} is computed. The next position is then uniformly sampled on the sphere $B(p, d)$. If the distance to the boundary is less than $\epsilon$, the current position is projected onto the boundary to obtain the sampled boundary value.

The WoS algorithm can also be understood through the mean value property of harmonic functions:
\begin{equation}
    u(x) = \frac{1}{\vert \partial B(x, r) \vert} \int_{\partial B(x, r)} u(y) \mathrm{d} y
\end{equation}
This recursive integration resembles the structure of the rendering equation in physically based rendering \cite{kajiya1986rendering}, and can be solved via single-sample Monte Carlo estimation.

\citet{sawhney2020monte} pioneered the application of WoS in geometry processing. PDEs are ubiquitous in this field, and traditional approaches often require extensive preprocessing and watertight meshes. \changed{Stochastic PDE solvers} offer \changed{an alternative} without discretization requirements. Subsequent works expanded this framework: \citet{sawhney2022grid} addressed spatially varying coefficients, \citet{sawhney2023walk} developed the Walk on Star method for Neumann boundary conditions, and \citet{miller2024walkin} generalized the walk on star algorithm to handle Robin conditions. Differentiable solvers for inverse problems were explored by \citet{miller2024differential, yu2024differential} and \citet{yilmazer2024solving}. \changed{Besides the WoS-type Monte Carlo solvers, \citet{sugimoto2023practical} introduced the walk-on-boundary method, providing an alternative for Monte Carlo geometry processing.}

\subsection{Variance Reduction Methods}

Beyond extending solver applicability, another major focus is accelerating convergence through variance reduction. \citet{sawhney2020monte} employed importance sampling of Green's functions and control variates to reduce variance in the WoS algorithm. Boundary value caching \cite{miller2023boundary} leveraged boundary integration to estimate solutions at evaluation points using cached samples from the boundary and domain. Mean value caching \cite{bakbouk2023mean} sampled domain points and evaluated solutions by utilizing the volumetric mean value property within the maximal sphere of each evaluation point. Reverse WoS \cite{qi2022bidirectional} proposed a bidirectional formulation, emitting random walking particles from both evaluation points and boundaries. More recently, \citet{huang2025path} adapted path-guiding methods from rendering to \changed{stochastic PDE solvers}, significantly improving cases with sparse source terms. Neural methods have also been explored: \citet{li2023neural} developed neural networks for caching estimation results, while \citet{li2024neural} employed neural networks to automate control variate selection. The works most related to our approach are \citet{hwang2015off} and \citet{czekanski2024walking}. \citet{hwang2015off} used the off-center formulation to accelerate the random walk. \citet{czekanski2024walking} leveraged the Poisson kernel to reuse samples from neighbors, but only \changed{Laplace} equations are handled.

Despite these advances, challenges remain. For instance, 3D diffusion problems with spatially varying source terms introduce high variance in stochastic PDE solvers. As shown in \changed{Figure \ref{fig:teaserfigure}}, the \changed{mean} value caching method \changed{generates some correlation artifacts}, especially at the positions close to the boundary. Even with the non-uniform sample used \changed{by \citet{bakbouk2023mean}}, some evaluation points close to the boundary can not get enough samples and lead to large errors. Benefiting from the global formulation, boundary value caching \cite{miller2023boundary} greatly reduces variance compared with \citet{sawhney2020monte}. But with very few samples and relatively short running time, boundary value caching generates results with relatively large errors.

\subsection{Sample Reuse in Rendering}

Monte Carlo rendering has been developed over decades, and many sample reuse strategies were proposed, such as \changed{virtual point lights \cite{keller1997instant, dachsbacher2014scalable}}, photon mapping \cite{jensen1996global}, and ReSTIR \cite{bitterli2020spatiotemporal, ouyang2021restir, lin2022generalized}. Many previous variance reduction techniques in \changed{stochastic PDE solvers} draw inspiration from rendering. For example, \citet{qi2022bidirectional} adapted bidirectional path tracing to WoS, while boundary value caching \cite{miller2023boundary}, inspired by virtual point lighting, employed boundary integration to cache values and derivatives for internal points.

The off-centered estimator shares some similarity with the rendering techniques, such as multiple importance sampling (MIS) \cite{veach1995optimally} and spatial sample reuse \cite{bitterli2020spatiotemporal}. To estimate an integral, MIS uses more than one distribution for importance sampling, and the off-centered estimator uses multiple integrals in different domains to estimate the value. The spatial part in ReSTIR \cite{bitterli2020spatiotemporal} reuses samples from neighboring pixels, and the off-centered estimator also reuses the data from neighboring evaluation points.

\begin{table}[h!]
    \centering
    \caption{Mathematical symbols and their definition.}
    \begin{tabular}{cr}
    \toprule
    \textbf{Symbol} & \textbf{Definition} \\
    \midrule
    $u$ & Solution of the boundary value problem \\
    $g$ & Absorption boundary term \\
    $h$ & Reflection boundary term \\
    $f$ & Source term \\
    $r_x$ & Distance from $x$ to boundary \\
    $B_x$ & Maximum ball $B(x, r_x)$ \\
    $\mathcal{D}$ & \changed{Second-order linear elliptic operator} \\
    $\vec{n}$ & \changed{Outward normal vector} \\
    $\Phi(r)$ & Fundamental solution \\
    $G^U$ & Green function in domain $U$ \\
    $I$ & Estimator \\
    $\hat{I}$ & Sample mean of estimator $I$\\
    \bottomrule
    \end{tabular}
    \label{tbl:notation}
\end{table}

\section{Method}

% In this section, we introduce our method of reusing samples in the Monte Carlo solvers. For vanilla Monte Carlo solvers, estimation at each point is completely uncorrelated. For each evaluation point $x$, the integration is done in $B_x$ and $\partial B_x$ independently. By the integration formula with Green's function, samples on $\partial B_x$ can be used to estimate solutions not only at the center $x$ but also anywhere else in the ball $B_x$. Since one evaluation point is usually covered by many maximal spheres of other evaluation points (like Figure \ref{fig:multiple_domain_integration}), we will have a variety of unbiased estimators for one evaluation point. Combining these estimators will greatly reduce the variance of stochastic solvers.

Generally, \changed{we consider the boundary value problem (\ref{equ:bvp2}) with Dirichlet boundary $\partial \Omega_D$ and Neumann boundary $\partial \Omega_N$,} and we assume that the centered Green's function of the \changed{second-order linear elliptic operator} $\mathcal{D}$ is known. The off-centered Green's function can at least be properly approximated.
\begin{equation}
\label{equ:bvp2}
\left\{
\begin{aligned}
    u(x) &= g(x), && x \in \partial \Omega_D, \\
    \frac{\partial u}{\partial \vec{n}} &= h(x), && x \in \partial \Omega_N, \\
    \mathcal{D} u (x) &= -f(x), && x \in \Omega.
\end{aligned}
\right.
\end{equation}
Our goal is to estimate the solution of the boundary value problem(BVP), described by a set of evaluation points $R = \{x_i\}_{i = 1}^N$ in $\Omega$. For $x, y$ in $R$, if $x$ is contained in $B_y$, we say that $y$ is a possible \textit{neighbor} of $x$. For an estimator $I$, we use $\hat{I}$ to represent the sample mean and $\mathrm{Var}(\hat{I})$ to denote sample variance. Notations are listed in table \ref{tbl:notation}.

\begin{figure}[h]
\centering
\includegraphics[width=0.8\linewidth]{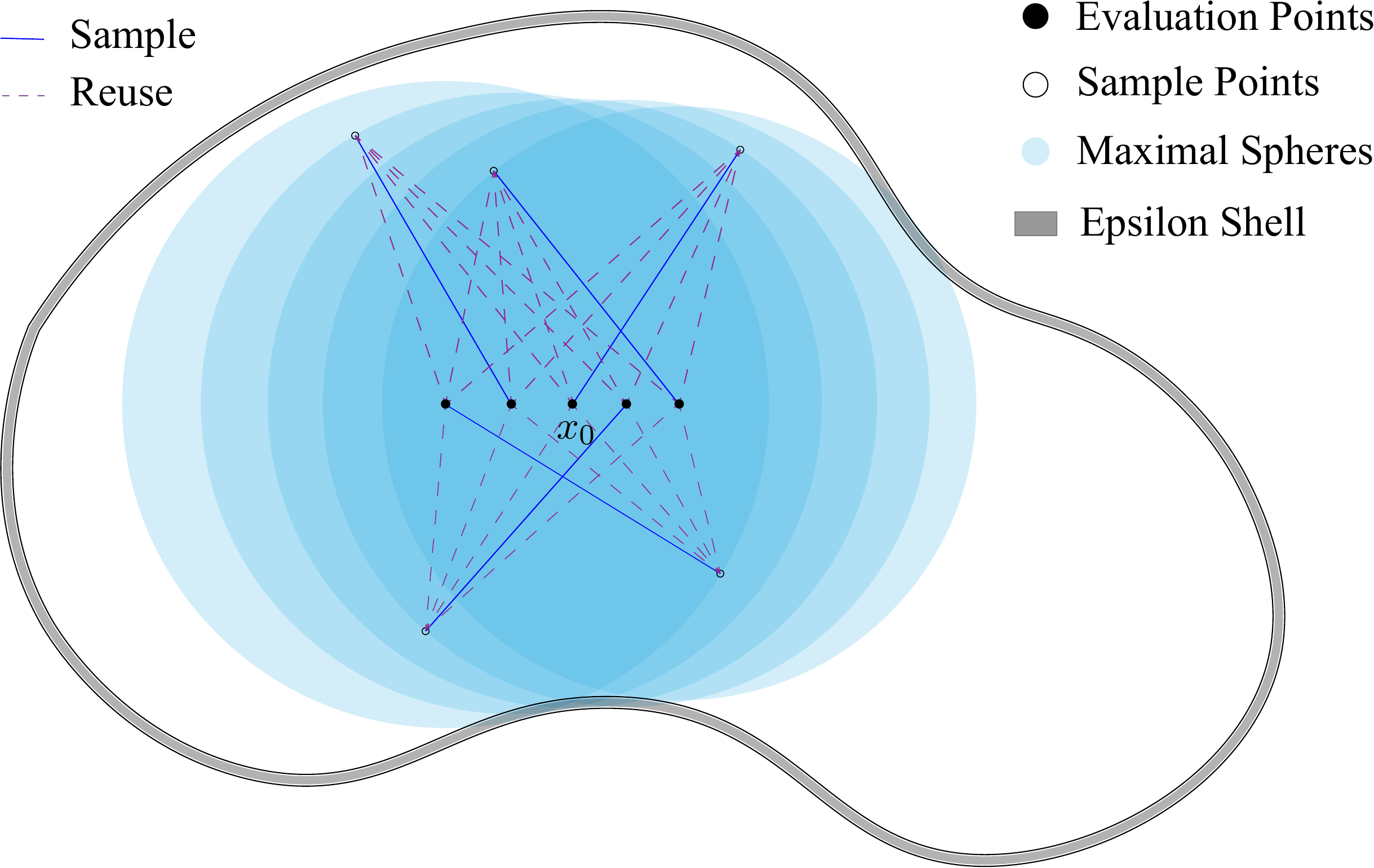}
\caption{Illustration of multiple domain integration for the Walk on Spheres algorithm. For a certain evaluation point $x_0$, the solution can be estimated using integrals over different domains—i.e., maximal spheres centered at various locations. The sampling stage and reuse stage are illustrated by solid and dashed lines, respectively.}
\label{fig:multiple_domain_integration}
\end{figure}

\subsection{Off-centered Estimator}

We state the off-centered estimator in a more general setting. From the PDE theory \cite{evans2022partial}, the solution $u(x)$ can be expressed as an integral over $B_x$ and its boundary $\partial B_x$, which we also call the centered estimator $I_{x, x}$:
\begin{equation}
u(x) = I_{x, x} = \int_{\partial B_x} u(y) P^{B_x}(x, y) \mathrm{d} S_y + \int_{B_x} f(y) G^{B_x}(x, y) \mathrm{d} y,
\end{equation}
where $P^{B_x}(x,y)$ is the Poisson kernel for the ball $B_x$, \changed{which acts as a weighting function that averages values of $u$ on the boundary $\partial B_x$ to determine $u$ at the interior point $x$. The function $G^{B_x}(x,y)$ is the Green's function for $B_x$, and $\mathrm{d}S_y$ is the differential area element on the sphere $\partial B_x$.} This leads to the classical Walk on Spheres (WoS) estimator \cite{sawhney2020monte}:

% \begin{mdframed}[style=mdfexample1, frametitle={Walk on Sphere Estimator $\bar{u}(x)$}]
% \begin{equation}
% \label{equ:wos}
% \begin{aligned}
%     \bar{u}(x) &= \frac{1}{N} \sum_{i = 1}^N \frac{\bar{u}(z_{x, i})P^{B_x}(x, z_{x, i})}{p_x(z_{x, i})} \\
%     &+ \frac{1}{M} \sum_{j = 1}^M \frac{f(w_{x, j}) G^{B_{x}}(x, w_{x, j})}{q_x(w_{x, j})}
% \end{aligned}
% \end{equation}
% $p_x(z)$ is a distribution on the sphere $\partial B_x$ and $q_{x}(z)$ is the pdf of some distribution in the ball $B_{x}$.
% \end{mdframed}

\newlength{\boxwidth}
\setlength{\boxwidth}{\dimexpr\linewidth-2\fboxsep-2\fboxrule\relax}

\noindent
\begin{minipage}{\linewidth}
  % Title
  \fcolorbox{white}{gray!20}{%
    \begin{minipage}{\boxwidth}
    \textbf{Walk on Sphere Estimator $\bar{u}(x)$}
    \end{minipage}%
  }

  % Content
  \fcolorbox{white}{gray!10}{%
    \begin{minipage}{\boxwidth}
    \begin{equation}
    \label{equ:wos}
    \begin{aligned}
        \bar{u}(x) &= \frac{1}{N} \sum_{i = 1}^N 
            \frac{\bar{u}(z_{x, i})P^{B_x}(x, z_{x, i})}{p_x(z_{x, i})} \\
            &+ \frac{1}{M} \sum_{j = 1}^M 
            \frac{f(w_{x, j}) G^{B_{x}}(x, w_{x, j})}{q_x(w_{x, j})}
    \end{aligned}
    \end{equation}

    $p_x(z)$ is a distribution on the sphere $\partial B_x$ and 
    $q_{x}(z)$ is the pdf of some distribution in the ball $B_{x}$.
    \end{minipage}%
  }
\end{minipage}
% \begin{examplebox}{Walk on Sphere Estimator $\bar{u}(x)$}
% \begin{equation}
% \label{equ:wos}
% \begin{aligned}
%     \bar{u}(x) &= \frac{1}{N} \sum_{i = 1}^N 
%        \frac{\bar{u}(z_{x, i}) P^{B_x}(x, z_{x, i})}{p_x(z_{x, i})} \\
%     &+ \frac{1}{M} \sum_{j = 1}^M 
%        \frac{f(w_{x, j}) G^{B_{x}}(x, w_{x, j})}{q_x(w_{x, j})}
% \end{aligned}
% \end{equation}

% $p_x(z)$ is a distribution on $\partial B_x$ and 
% $q_x(z)$ is the pdf of some distribution in $B_x$.
% \end{examplebox}

When $x$ lies inside $B_y$, the maximal ball centered at another evaluation point $y$, $u(x)$ can also be expressed via alternative integrals. We denote these off-centered estimators as $I_{x, y}$, which means estimating the solution $u(x)$ via data from $\partial B_y$ and $B_y$:
\begin{equation}
\label{equ:offcenter}
u(x) = I_{x, y} = \int_{\partial B_y} u(z) P^{B_y}(x, z) \mathrm{d} S_z + \int_{B_y} f(z) G^{B_y}(x, z)  \mathrm{d} z.
\end{equation}

The off-centered formulation allows us to construct multiple estimators $I_{x, y}$ for each evaluation point $x$. As shown in Figure \ref{fig:multiple_domain_integration}, We choose a set of neighboring evaluation points $H_x = \{y : \Vert x - y \Vert < r_y\}$, and form a weighted combination:
\begin{equation}
\label{equ:wmdi_short}
\hat{u}(x) = \sum_{y \in H_x} \lambda_{\changed{x, y}} \hat{I}_{x, y},
\end{equation}
where the weights $\lambda_{\changed{x, y}}$ satisfy $\sum_{y \in H_x} \lambda_{\changed{x, y}} = 1$. This leads to the multiple domain estimator $\hat{u}$. \changed{We use $\bar{u}$ for estimators without sample reuse and use $\hat{u}$ for the combined estimators.}
% \begin{mdframed}[style=mdfexample1, frametitle={Multiple Domain Estimator $\hat{u}(x)$}]
% \changedd{We first compute the off-centered estimators}
% \begin{equation}
% \label{equ:mdi}
% \begin{aligned}
%     \changedd{\hat{I}_{x, y}} &= \frac{1}{\changed{N}} \sum_{i = 1}^N  \frac{\bar{u}(z_{y, i}) P^{B_y} (x, z_{y, i}) }{p_{\changed{x, y}}(z_{y, i})}\\
%     &+ \frac{1}{\changed{M}} \sum_{j = 1}^M  \frac{f(w_{y, j}) G^{B_{y}}(x, w_{y, j})}{q_{x, y}(w_{y, j})}. \\
% \end{aligned}
% \end{equation}
% \changedd{Then combine them with equation \ref{equ:wmdi_short} to get $\hat{u}(x)$.}

% $p_x(z)$ is a distribution on the sphere $\partial B_x$ and $q_{x, y}(z)$ is the pdf of some distribution in the ball $B_{y}$ specially adapted for $x$. We introduce the sampling details in section \ref{sec:sampling}.
% \end{mdframed}

\noindent
\begin{minipage}{\linewidth}
  % Title bar
  \fcolorbox{white}{gray!20}{%
    \begin{minipage}{\boxwidth}
    \textbf{Multiple Domain Estimator $\hat{u}(x)$}
    \end{minipage}%
  }

  % Content box
  \fcolorbox{white}{gray!10}{%
    \begin{minipage}{\boxwidth}
    \changedd{We first compute the off-centered estimators}
    \begin{equation}
    \label{equ:mdi}
    \begin{aligned}
        \changedd{\hat{I}_{x, y}} &= \frac{1}{\changed{N}} 
        \sum_{i = 1}^N  
        \frac{\bar{u}(z_{y, i}) P^{B_y} (x, z_{y, i}) }{p_{\changed{x, y}}(z_{y, i})}\\
        &+ \frac{1}{\changed{M}} 
        \sum_{j = 1}^M  
        \frac{f(w_{y, j}) G^{B_{y}}(x, w_{y, j})}{q_{x, y}(w_{y, j})}. \\
    \end{aligned}
    \end{equation}
    \changedd{Then combine them with equation \ref{equ:wmdi_short} to get $\hat{u}(x)$.}

    $p_x(z)$ is a distribution on the sphere $\partial B_x$ and 
    $q_{x, y}(z)$ is the pdf of some distribution in the ball $B_{y}$ 
    specially adapted for $x$. We introduce the sampling details 
    in section \ref{sec:sampling}.
    \end{minipage}%
  }
\end{minipage}

% \begin{examplebox}{Multiple Domain Estimator $\hat{u}(x)$}
% \changedd{We first compute the off-centered estimators}
% \begin{equation}
% \label{equ:mdi}
% \begin{aligned}
%     \changedd{\hat{I}_{x, y}} &= \frac{1}{\changed{N}} 
%         \sum_{i = 1}^N \frac{\bar{u}(z_{y, i}) P^{B_y}(x, z_{y, i})}{p_{\changed{x, y}}(z_{y, i})} \\
%     &+ \frac{1}{\changed{M}} 
%         \sum_{j = 1}^M \frac{f(w_{y, j}) G^{B_{y}}(x, w_{y, j})}{q_{x, y}(w_{y, j})}.
% \end{aligned}
% \end{equation}

% \changedd{Then combine them with equation \ref{equ:wmdi_short} to get $\hat{u}(x)$.}

% $p_x(z)$ is a distribution on the sphere $\partial B_x$ and 
% $q_{x, y}(z)$ is the pdf of some distribution in the ball $B_{y}$ specially adapted for $x$. 
% We introduce the sampling details in section \ref{sec:sampling}.
% \end{examplebox}

The first problem in estimator \ref{equ:mdi} is the evaluation of $G^{B_y}(x, z)$. In equation \ref{equ:mdi}, we densely use the off-centered Green's function $G^{B_y}(x, z)$. For the Poisson equation, the Green's function can be exactly computed via the mirror method:
\begin{equation}
\label{equ:green}
G^{B_y}(x, z) = \Phi(|x - z|) - \Phi\left(\frac{|x - y|}{r_y} |x^* - z|\right).
\end{equation}
For the \changed{screened Poisson equation ($\mathcal{D} = \Delta - \sigma$, $\sigma > 0$ is the screening coefficient)}, the exact Poisson kernel and Green's function are under a series form, and we use the approximation provided \changed{by \citet{sawhney2022grid}}. Different from the Poisson case, the approximated Green's function leads to additional bias.

The second term to be determined is the weighting coefficients $\lambda_{x, y}$. \citet{czekanski2024walking} considered the simplest setting where $\mathcal{D} = \Delta$ and $f = 0$. In the Laplacian case, they proved that the variance of the estimator $I_{x, y}$ can be bounded using properties of the Poisson kernel:
\begin{equation}
\mathrm{Var}(I_{x, y}) \leq \frac{\left(1 - r^{-2} \Vert x - y \Vert^2\right)^2}{\left(1 - r^{-1} \Vert x - y \Vert\right)^{2d}} M^2,
\end{equation}
where $d$ is the spatial dimension, $M = \frac{1}{2} \max_{x \in \partial \Omega} |\changed{g(x)}|$,  \changed{r is the radius of $B_y$}, and the pre-factor before $M^2$ is an upper bound on the Poisson kernel $P^y(x, z)$. Based on this bound, they proposed the following weight for variance reduction:
\begin{equation}
\label{equ:poisson_bound}
\lambda_{x, y} \propto \frac{\left(1 - r^{-1} \Vert x - y \Vert\right)^{2d}}{\left(1 - r^{-2} \Vert x - y \Vert^2\right)^2}.
\end{equation}
For the Laplace equation, the Poisson bound weight performs well, but for more complex equations, this value can not properly bound the variance of estimators. 

\subsection{Statistical Weighting}
\label{sec:weighting}

\begin{figure}[h]
    \centering
    \includegraphics[width=0.9\linewidth]{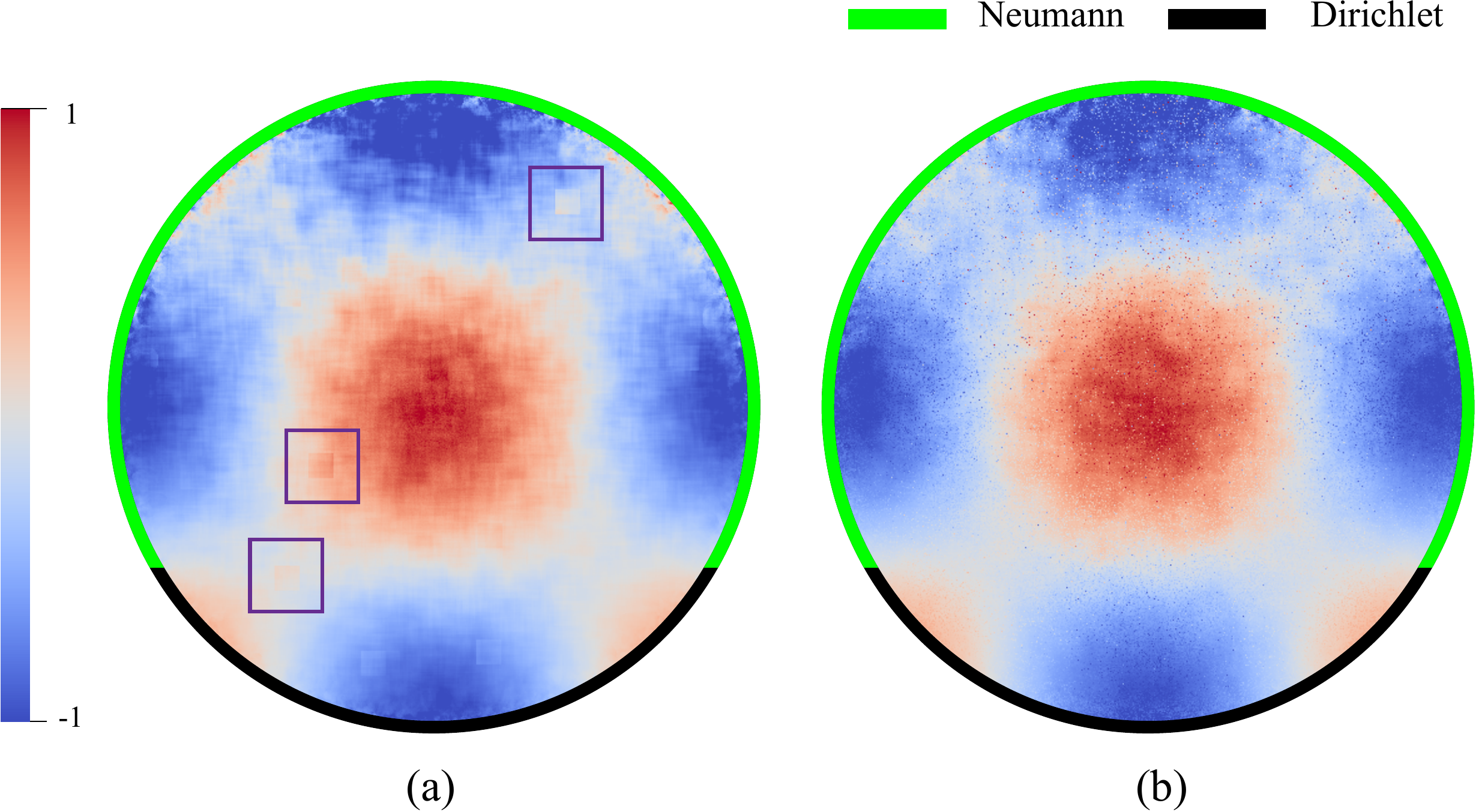}
    \caption{Outlier values in off-centered estimators lead to remarkable artifacts in visual results. (a): off-centered estimator \changed{with the uniform weighting}, extreme values affect all evaluation points that reuse them. Artifacts caused by outliers are marked by purple boxes (b): Our statistical weighting strategy can significantly reduce these artifacts. }
    \label{fig:artifact}
\end{figure}

In this section, we describe our choice of weighting coefficients in estimator~\ref{equ:mdi}. We adopt a filter-based denoising technique to determine proper coefficients, addressing the following two issues:

\begin{itemize}
\item The off-centered estimator introduces correlations between evaluation points. A single outlier \changed{with large error} can contaminate multiple evaluations that reuse it, causing both visual artifacts and numerical errors (see Figure \ref{fig:artifact}).
\item When Green's function is approximated, off-centered estimators are biased. To mitigate this, we must identify low-quality estimators and apply a bias-variance strategy (see Figure \ref{fig:graphical_abstract}).
\end{itemize}

Recently, \citet{sakai2024statistical} proposed a statistical filtering framework for Monte Carlo rendering. Their key idea is to evaluate the similarity between nearby pixels using an efficient pairwise statistical test. Following this approach, we define the weight in Equation~(\ref{equ:wmdi_short}) as:

\begin{equation}
    \lambda_{\changed{x, y}} = \frac{m_{xy}}{\sum_{y \in H_x} m_{xy}},
\end{equation}
where $m_{xy}$ is determined based on the statistics $w^*$:

\begin{equation}
    \label{equ:wstar}
    w^* = \frac{(\hat{I}_{x, x} - \hat{I}_{x, y})^2 + \mathrm{Var}(\hat{I}_{x, y})}{(\hat{I}_{x, x} - \hat{I}_{x, y})^2 + \mathrm{Var}(\hat{I}_{x, y}) + \mathrm{Var}(\hat{I}_{x, x})}
\end{equation}
If $1 - w^*$ exceeds a predefined threshold $\gamma$, we set $m_{xy} = 1$; otherwise, we set $m_{xy} = 0$. \changed{The variance of $\hat{I}_{x,y}$ is evaluated on the fly using the stored sample mean and mean of squares.}

Intuitively, this approach treats the center estimator $I_{x, x}$ as a reference for similarity evaluation. If an off-centered estimator $I_{x, y}$ is sufficiently similar to $I_{x, x}$, it is incorporated into the computation. Conversely, if a sample from $y$ results in an outlier, it is excluded from reuse in neighboring evaluations, thereby reducing the impact of \changed{estimators with large error}. As shown in Figure \ref{fig:artifact}, using this sampling strategy can remarkably reduce the correlation artifacts in the off-centered estimator. In Figure \ref{fig:artifact} (b), there are many noisy points, indicating that outlier estimation happens. Proposed weighting strategy prevents other evaluation points from reusing the extreme data and hence reduces the artifact in Figure \ref{fig:artifact} (a).

\changed{With statistical weighting, the estimator in equation \ref{equ:mdi} is no longer unbiased even with all $I_{x, y}$ unbiased due to the weights depending on the estimates themselves (we do not consider the bias brought by $\epsilon$-shell for convenience). Although biased, the convergence can be guaranteed since the error of the combined estimator is bounded by the largest error of the independent estimators:
\begin{equation}
    \label{equ:error}
    \vert\hat{u}(x) - u(x) \vert \leq \sum_{y \in H_x} \lambda_{x, y} \vert \hat{I}_{x, y} - u(x) \vert \leq \max_{y \in H_x} \vert \hat{I}_{x, y} - u(x) \vert.
\end{equation}
When all estimators $I_{x, y}$ are unbiased, the right-hand side converges to zero and hence $\hat{u}(x)$ will converge to $u(x)$ as $N \to \infty$.}

A potential question is why we select $\hat{I}_{x, x}$ as the reference instead of $\hat{I}_{x, y}$ with the minimal variance. The reason lies in the structure of Equation~(\ref{equ:wstar}): similarity depends not only on variance but also on the mean. An estimator with minimal variance may still have a mean far from the expected value, potentially introducing a \changed{larger error}.

Another benefit provided by this weighting approach is the bias-variance trade-off. In the screened Poisson case, we use an approximation of the off-centered Green's function, which will introduce additional bias to the off-centered estimators $I_{x, y}$. Note that if $1 - w^* > \gamma$, $I_{x, y}$ passes the similarity test \changed{and the distance between} $\hat{I}_{x, x}$ and $\hat{I}_{x, y}$ is controlled by $\mathrm{Var}(\hat{I}_{x, x})$
\begin{equation}
    (\hat{I}_{x, x} - \hat{I}_{x, y})^2 < (\frac{1}{\gamma} - 1) \mathrm{Var}(\hat{I}_{x, x}) - \mathrm{Var}(\hat{I}_{x, y}) \leq (\frac{1}{\gamma} - 1) \mathrm{Var}(\hat{I}_{x, x}).
\end{equation}
When the variance of the center estimate $\hat{I}_{x, x}$ is high, the method allows some biased off-centered estimators to be included to reduce overall variance. As more samples are collected, the variance of the center estimator decreases linearly, and those biased estimators are gradually excluded. Hence, the described weighting strategy can filter out the highly biased estimators $I_{x, y}$, and we can tune $\gamma$ to \changed{make a bias–variance trade-off.}

\subsection{Generalization}

In this section, we demonstrate how our multiple-domain estimator can be generalized to handle diverse boundary conditions, integrate with different pointwise estimators, and support gradient estimation. These extensions preserve the core variance-reduction benefits of our approach while broadening its applicability to a wider class of problems.

\begin{figure}
    \centering
    \includegraphics[width=\linewidth]{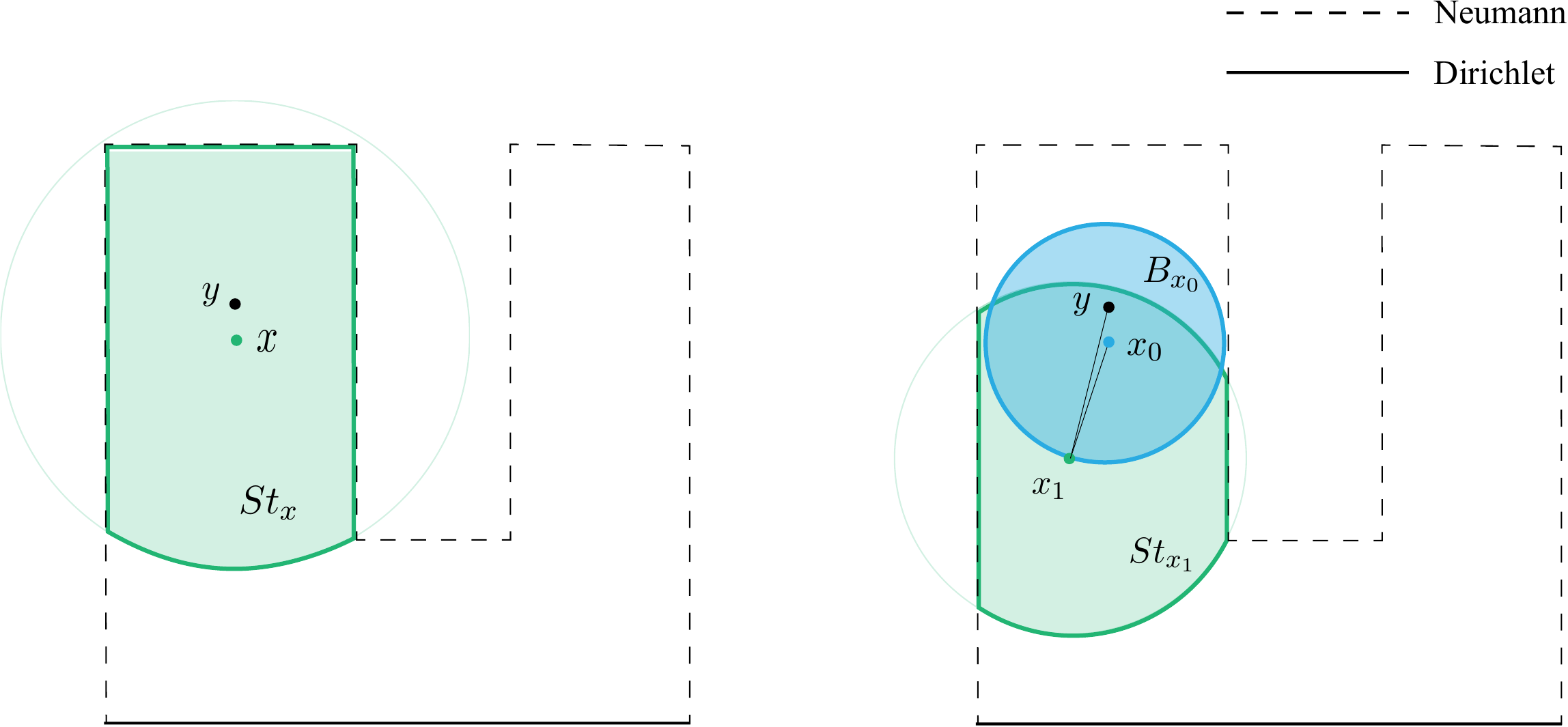}
    \caption{Left: the first step of the standard walk-on-star algorithm. Right: our method extends walk-on-star estimators by introducing an additional step in the random walk. Blue spheres represent added integration domains, and the green region shows the second step of the walk-on-star algorithm for a sample in one of these domains. The estimated value $\hat{u}(x_1)$ is used to estimate both center position $x_0$ and off-centered position $y$.}
    \label{fig:extra_step}
\end{figure}

% For mixed Dirichlet and Neumann boundary value problems, the walk-on-star (WoSt) algorithm~\cite{sawhney2023walk} replaces balls with star-shaped domains to simulate reflecting Brownian motion. The solution \( u(x) \) is expressed as integrals over \( S_x \) and \( \partial S_x \), where \( S_x \) is a star-shaped domain. However, since the Green's function for an arbitrary star-shaped domain is generally unavailable, directly applying our sample reuse strategy in this setting is infeasible. To address this, we introduce an additional step in the WoSt random walk. Specifically, we replace the largest star-shaped domain \( S_x \) with a maximal inscribed ball \( B_x \). By doing so, we enable sample reuse from the added balls that cover \( S_x \), making our estimator compatible with WoSt-based boundary handling.

\textit{Boundary conditions.} Walk on star(WoSt) algorithm~\citet{sawhney2023walk} efficiently handles the mixed Dirichlet and Neumann boundary conditions. In the WoSt algorithm, the solution $u(x)$ is estimated by data in a star-shaped domain $St_x$. Directly reusing the samples on $\partial St_x$ to estimate other $y \in St_x$ is difficult. \changed{For instance, $St_x$ is not guaranteed to be a star-shaped domain with respect to $y$.} Hence, for the mixed domain boundary problems, we replace the star-shaped domain $St_x$ with the inscribed ball $B_x$ (Figure \ref{fig:extra_step}). Then, from samples on $x_1 \in B_x$, we estimate solution $u(x_1)$ with the Walk on Star estimator and reuse these samples via the off-centered scheme.

\textit{Gradient estimation.}  
\citet{sawhney2020monte} introduced a pointwise estimator for the gradient \changed{\( \nabla u(x) \)} of the solution using:
\begin{equation}
    \changed{\nabla u(x)} = \frac{1}{\vert \partial B_x \vert} \int_{\partial B_x} u(y) \nu(y) \, \mathrm{d}y + \int_{B_x} f(y)\nabla_x G(x,y) \, \mathrm{d}y,
\end{equation}
where \( \nu(y) \) is the outward normal at \( y \in \partial B_x \). This can be extended to an off-center formulation when \( x \in B_y \):
\begin{equation}
    \changed{\nabla u(x)} = \frac{1}{\vert \partial B_y \vert} \int_{\partial B_y} u(z) \nabla_x P(x, z) \, \mathrm{d}z + \int_{B_y} f(z) \nabla_x G(x, z) \, \mathrm{d}z.
\end{equation}
Based on this, we also derive an off-center estimator for the gradient estimation. 

% \begin{mdframed}[style=mdfexample1, frametitle={Gradient Multiple-Domain Estimator \changed{$\widehat{\nabla\! u}$}}]
% \begin{equation}
% \begin{aligned}
%     \changed{\widehat{\nabla\! u}(x)} &= \frac{1}{N} \sum_{i = 1}^N \sum_{y \in H_x} \lambda_{x, y} \frac{\bar{u}(z_{y, i}) \, \nabla_x P^{B_{\changed{y}}}(x, z_{y, i})}{p_{\changed{x, y}}(z_{y, i})} \\
%     &\quad + \frac{1}{\changed{M}} \sum_{j = 1}^M \sum_{y \in H_x} \lambda_{x, y} \, \frac{f(w_{y, j}) \, \nabla_x G^{B_y}(x, w_{y, j})}{q_{x, y}(w_{y, j})}.
% \end{aligned}
% \end{equation}
% \end{mdframed}

\noindent
\begin{minipage}{\linewidth}
  % Title bar
  \fcolorbox{white}{gray!20}{%
    \begin{minipage}{\boxwidth}
    \textbf{Gradient Multiple-Domain Estimator \changed{$\widehat{\nabla\! u}$}}
    \end{minipage}%
  }

  % Content box
  \fcolorbox{white}{gray!10}{%
    \begin{minipage}{\boxwidth}
    \begin{equation}
    \begin{aligned}
        \changed{\widehat{\nabla\! u}(x)} &= \frac{1}{N} \sum_{i = 1}^N \sum_{y \in H_x} 
        \lambda_{x, y} \frac{\bar{u}(z_{y, i}) \, \nabla_x P^{B_{\changed{y}}}(x, z_{y, i})}{p_{\changed{x, y}}(z_{y, i})} \\
        &\quad + \frac{1}{\changed{M}} \sum_{j = 1}^M \sum_{y \in H_x} 
        \lambda_{x, y} \, \frac{f(w_{y, j}) \, \nabla_x G^{B_y}(x, w_{y, j})}{q_{x, y}(w_{y, j})}.
    \end{aligned}
    \end{equation}
    \end{minipage}%
  }
\end{minipage}

% \begin{examplebox}{Gradient Multiple-Domain Estimator \changed{$\widehat{\nabla\! u}$}}
% \begin{equation}
% \begin{aligned}
%     \changed{\widehat{\nabla\! u}(x)} &= 
%         \frac{1}{N} \sum_{i = 1}^N \sum_{y \in H_x} 
%         \lambda_{x, y} \frac{\bar{u}(z_{y, i}) \, \nabla_x P^{B_{\changed{y}}}(x, z_{y, i})}
%         {p_{\changed{x, y}}(z_{y, i})} \\
%     &\quad + \frac{1}{\changed{M}} \sum_{j = 1}^M \sum_{y \in H_x} 
%         \lambda_{x, y} \, \frac{f(w_{y, j}) \, \nabla_x G^{B_y}(x, w_{y, j})}
%         {q_{x, y}(w_{y, j})}.
% \end{aligned}
% \end{equation}
% \end{examplebox}

The weighting coefficients $\lambda_{\changed{x, y}}$ \changed{are determined by the same strategy in section \ref{sec:weighting}.}

% \section{Implementation Details}

% In this section, we state the implementation of our stochastic solver system. We list the pseudo-code in algorithm \ref{alg:main} and explain details in subsections.

% \subsection{Neighbor Selection}

% \subsection{Importance Sampling}

\begin{algorithm}
\caption{Solve BVP with Proposed Method}
\label{alg:main}
\begin{algorithmic}[1]
\STATE \textbf{Input:} evaluation points $\{x_i\}_{1 \leq i \leq N}$, samples per pixel $S$, threshold $t$.
\STATE \textbf{Output:} estimation of BVP solution at evaluation points $\{u_i\}$.

\STATE \textcolor{blue}{\# Initialization: Get maximum spheres and neighbors}
\FOR{\textbf{parallel} $i = 1$ to $N$}
    \STATE $r_{x_i} \gets \text{DistanceQuery}(x_i)$
    \STATE $H_{x_i} \gets \text{SelectNeighbors}(x_i)$
\ENDFOR

\FOR{$k = 1$ to $S$}
    \STATE \textcolor{blue}{\# First stage: pointwise estimation (solid lines in Figure \ref{fig:multiple_domain_integration})}
    \FOR{\textbf{parallel} $i = 1$ to $N$}
        \STATE Sample $z_{x_i, k}$ on the sphere $B_{x_i}$
        \STATE Estimate $\bar{u}(z_{x_i, k})$ with pointwise estimator
    \ENDFOR

    \STATE \textcolor{blue}{\# Second stage: local sample reuse (dashed lines in Figure \ref{fig:multiple_domain_integration})}
    \FOR{\textbf{parallel} $i = 1$ to $N$}
        \STATE $\hat{u}_i \gets 0$, $w \gets 0$
        \FOR{$y$ in $H_{x_i}$}
            \STATE Compute $\hat{I}_{x_i, y}$ by equation (\ref{equ:mdi})
            \STATE Update the statistics of $\hat{I}_{x_i, y}$
            \STATE Compute $w^*$ by equation (\ref{equ:wstar})
            \STATE $w_y \gets 1$ if $1 - w^* > t$ otherwise $0$
            \STATE $\hat{u}_i \gets \hat{u}_i + w_y \hat{I}_{x_i, y}$
            \STATE $w \gets w + w_y$
        \ENDFOR
        \STATE $\hat{u}_i \gets \hat{u}_i / w$
        \STATE $u_i \gets (u_i (k - 1) + \hat{u}_i) / k$
    \ENDFOR
\ENDFOR

\STATE \textbf{Return} $\{u_{i}\}$
\end{algorithmic}
\end{algorithm}

\section{Implementation}
\label{sec:detail}

\subsection{Neighbor Selection}

% Suppose each estimation $\hat{I}_{x, y}$ is computed using $n$ samples. Note that different off-centered estimators $I_{x, y}$ are independent, then the expected squared error of the final estimator is:
% \begin{equation}
% \label{equ:bias_variance}
% \begin{aligned}
%     &\mathbb{E} \left(\sum_{y \in H_x} \lambda_{x, y} \hat{I}_{x, y} - u(x)\right)^2 \\ 
%     &= \mathbb{E}\left(\sum_{y\in H_x} \lambda_{x, y} \left(\hat{I}_{x, y} - \mathbb{E}(I_{x, y}) + \mathbb{E}(I_{x, y}) - u(x)\right)\right)^2 \\
%     &\leq \frac{1}{n H} \max_{y \in H_x} \mathrm{Var}(I_{x, y}) + \frac{1}{H^2} \sum_{y\in H_x}\left(\mathbb{E}(I_{x, y}) - u(x)\right)^2.
% \end{aligned}
% \end{equation}

\changed{Not all neighbors contribute to the variance reduction, and including estimators with large variance may increase the error. As shown in equation \ref{equ:error}, the largest error of the independent estimators bounds the error. Hence, we reject those estimators $I_{x, y}$ with large variance before the sampling stage. Although the statistical weighting can reject estimators with relatively large errors and sample variance, those estimators with large expected variance can still be harmful to variance reduction.} Empirically, if $x$ is closer to the boundary of $\partial B_y$, the estimator $I_{x, y}$ has larger variance, and we reject them before the sampling stage. We select the neighbor $H_x$ with the following equation
\begin{equation}
    H_x = \{y \mid d(x, y) < \min (\alpha r_y, \beta)\},
\end{equation}
where $\alpha \in (0, 1)$ and $\beta > 0$ are hyperparameters. \changed{Empirically, we use $\alpha = 0.5$ and $\beta = 10$ in our experiment.}

\changed{When the evaluation points are a dense grid ${x_{i, j}}$, we simply find neighbors by the indices, which is slightly faster and model size agnostic. This strategy is formally described by
\begin{equation}
    H_{x_{i, j}} = \{x_{p, q} \mid d(x_{i, j}, x_{p, q}) < \alpha r_{x_{p,q}},  \max(|p - i|, |q -j|) < \beta\}.
\end{equation}
For 3D grid $x_{i, j, k}$, we use similar approach described by $\max(|p - i|, |q -j|, |r - k|) < \beta$.}

\subsection{Importance Sampling}
\label{sec:sampling}

Care must be taken when choosing the sampling distribution $p_{x, y}(z)$ in equation \ref{equ:mdi}. For the Laplace operator $\mathcal{D}u = \Delta u$, the Green’s function in $B_y$ can be expressed using the method of images as shown in equation \ref{equ:green}. \changed{Note that in equation \ref{equ:green},} $G^{B_y}(x, z)$ becomes singular when $z \to x$. If $z$ is sampled naively from $G^{B_y}(y, z)$—as in the pointwise estimator—the value $G^{B_y}(x, z)$ may be large while the probability density $p_y(z)$ is small, resulting in high-variance outliers. The Green's function for the \changed{screened Poisson equation shared a similar issue}.

\changed{We sample the source term with a two-stage method:
\begin{enumerate}
    \item Sample ray with direction \changedd{$\vec{v}$} from \(\vec{x}\), intersect with the ball \(B_y\).  
    \item Sample \(\vec{w}=\vec{x}+u\vec{v}\) from \(u\sim G^{B(x, r_y + |x-y|)}\). If \(\vec{w}\) is not contained in \(B_y\), ignore source contribution. 
\end{enumerate}}
For the screened Poisson equation, the same sampling strategy is applied to the source term.

\section{Evaluation}

\subsection{Setup}

In this section, we show our results on solving a variety of 3D boundary value problems and compare the proposed method with vanilla WoS solver \cite{sawhney2020monte}, mean value caching \cite{bakbouk2023mean}, and boundary value caching \cite{miller2023boundary}.

We implemented the proposed algorithm in C++ and listed pseudo code in algorithm \ref{alg:main}. Geometric primitives are powered by the \changed{FCPW library \cite{FCPW}.} \changed{Experiments were run on a PC with an Intel i5-12500 processor, 32 GB RAM, and only CPU parallelization.} For the mean value caching algorithm, we use the non-uniform variant \changed{with parameters used by \citet{bakbouk2023mean}} since we found in practice that the evaluation points near the boundary commonly have \changed{larger error}. When there are no samples in the maximal ball of some evaluation point, we add a sample at the center and estimate it with the standard WoS or WoSt algorithm.

\subsection{Evaluation Problems}

\begin{figure}[h]
    \centering
    \includegraphics[width=\linewidth]{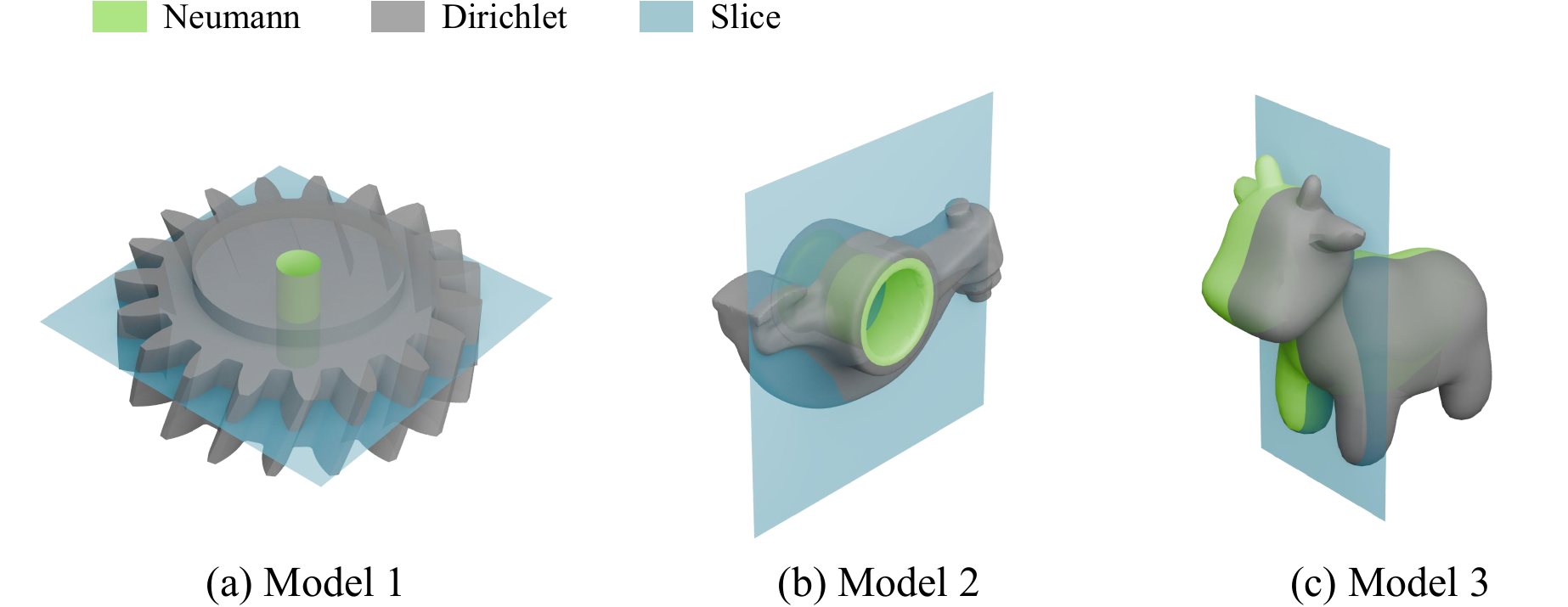}
    \caption{Boundary geometry, division of different boundary conditions, and evaluation slices in our experimentation. \changed{For the experimentation of pure Dirichlet problems, we use all Dirichlet boundary conditions.}}
    \label{fig:models}  
\end{figure}

We evaluate the performance of different solvers on a variety of problems. For the numerical data, we use a known ground truth solution of the form:
\begin{equation}
    \label{equ:poisson1}
    u(x, y, z) = \sin(\omega x) \sin(\omega y) \sin(\omega z),
\end{equation}
with the corresponding boundary and source terms, which allows for the precise error measurements and convergence analysis.
For the boundary geometry, we use different models and boundary conditions as shown in Figure \ref{fig:models}. To illustrate results visually, we estimate the solution on a 2D slice with $512 \times 512$ resolution, shown as the green slice in Figure \ref{fig:models}.

\subsection{Compare with Simple Off-centered Estimator}

\begin{figure}
    \centering
    \includegraphics[width=\linewidth]{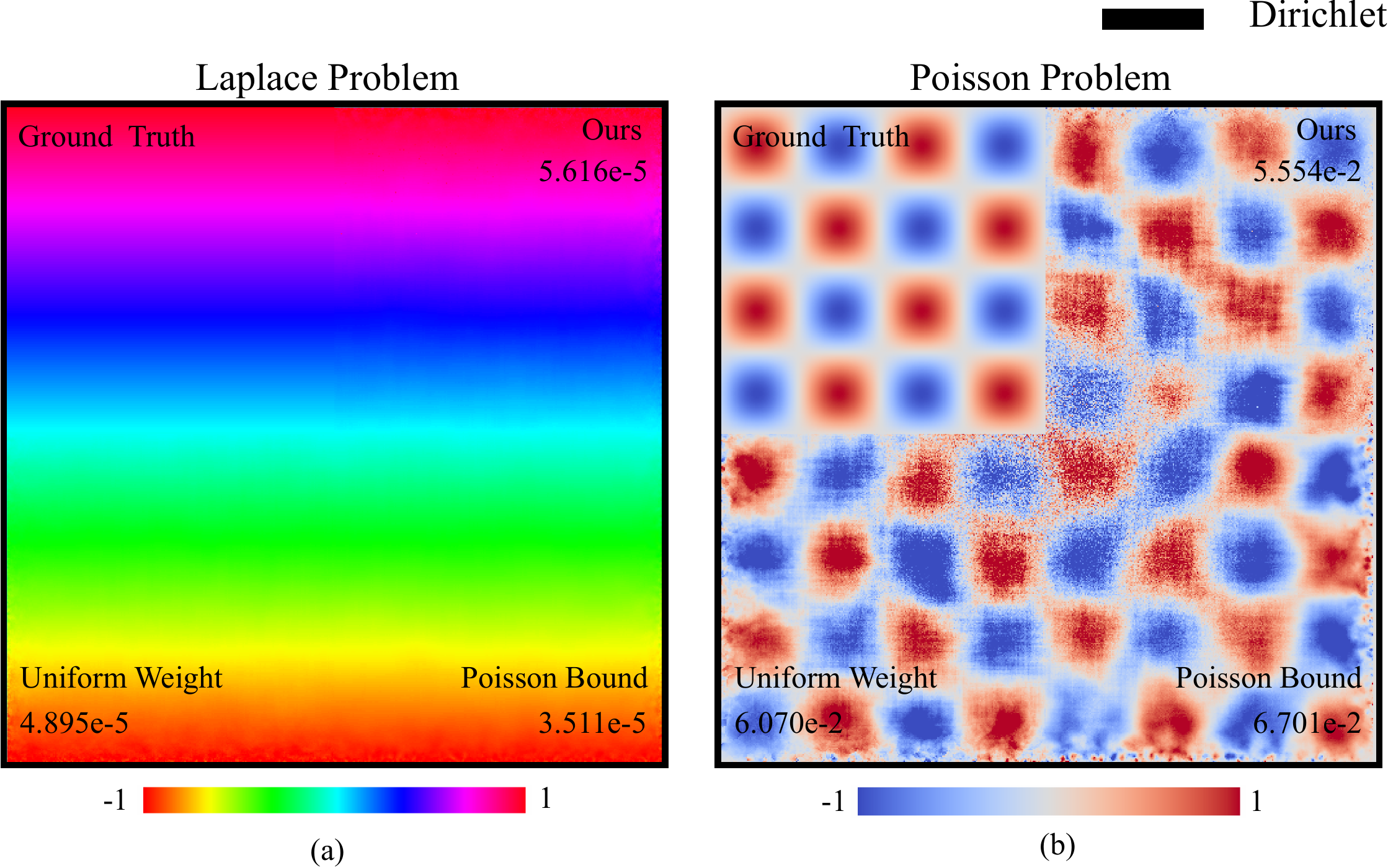}
    \caption{Result and MSE of different weighting methods in the simple 2D domain. \textbf{(a)}: For the simplest Laplace problem, our method has higher variance than the method by \citet{czekanski2024walking}. \textbf{(b)}: For the Poisson equations, the motivation of the Poisson bound weight no longer holds.} 
    \label{fig:2d_results}
\end{figure}

\begin{figure*}
    \centering
    \subfigbottomskip=2pt
    \subfigcapskip=-5pt
        \subfigure[]{
        \includegraphics[width=0.31\linewidth]{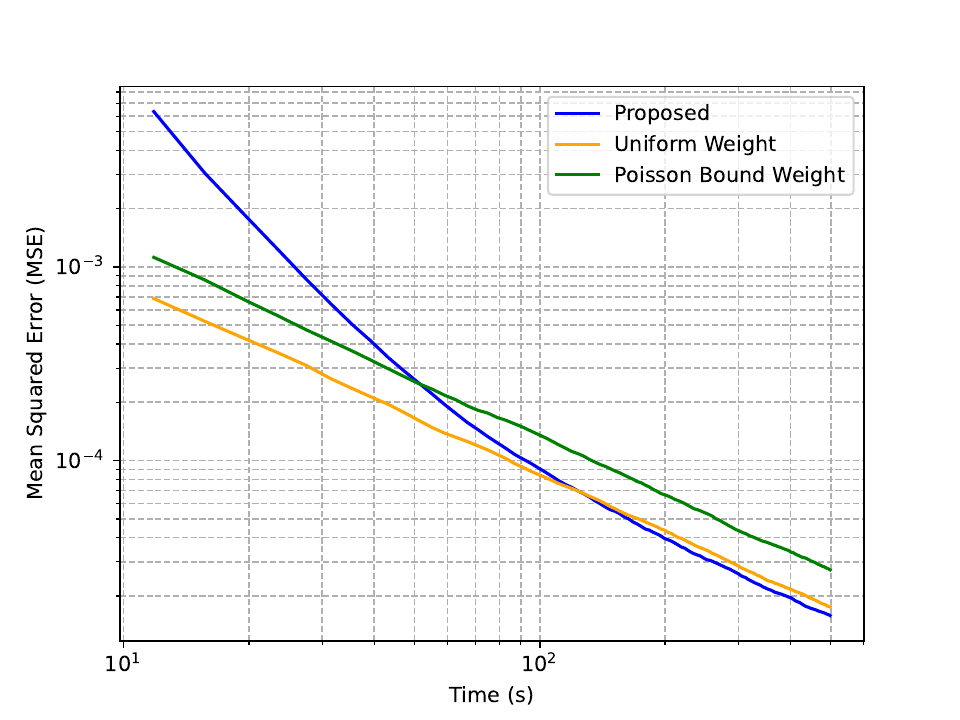}}
    \subfigure[]{
        \includegraphics[width=0.31\linewidth]{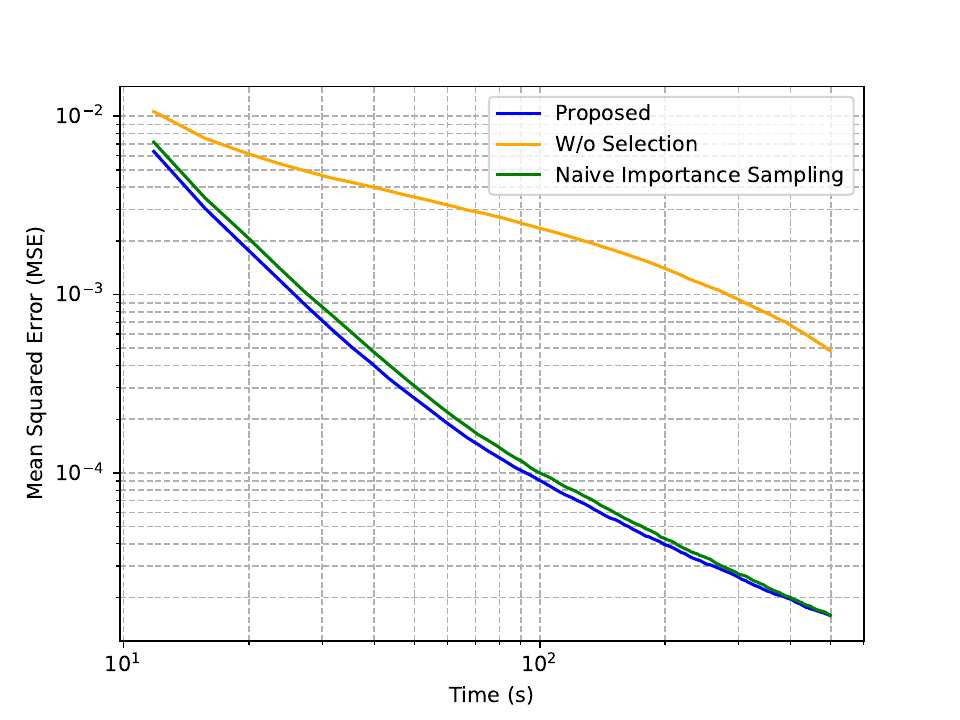}}
    \subfigure[]{
        \includegraphics[width=0.31\linewidth]{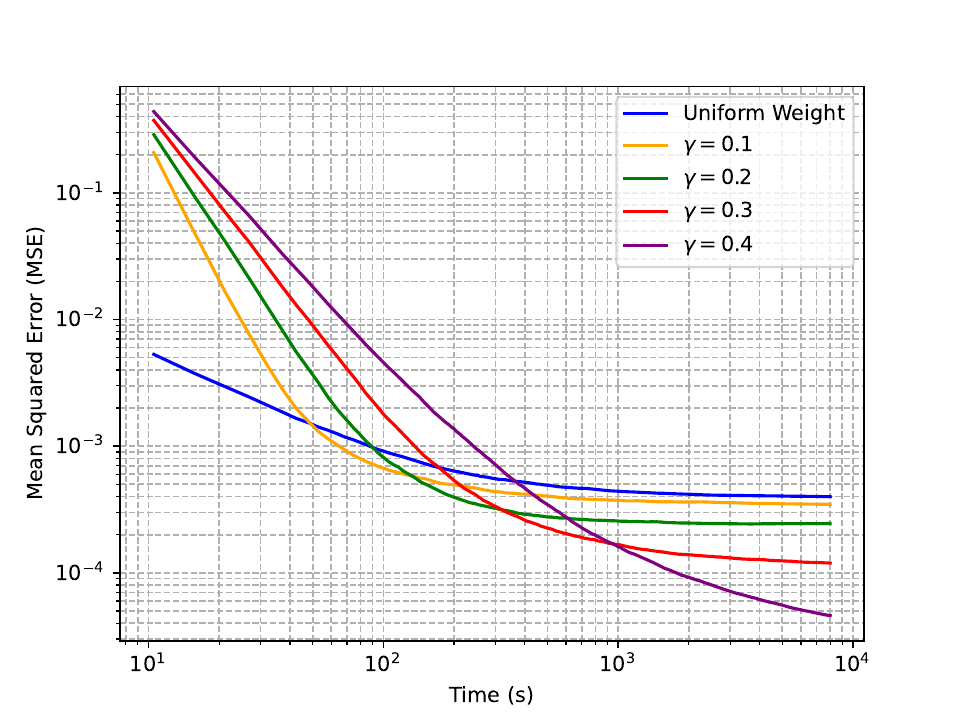}}\\
    \caption{(a): Comparison with the weighting strategy used \changed{by \citet{czekanski2024walking}}. The Poisson kernel bound weight is not effective and has brought negative effects. (b) Ablation result of the sampling strategy and neighbor selection strategy. Our sampling strategy slightly improved the MSE result, and without the neighbor selection, the error would be remarkably large. (c): Comparison of uniform weight and different $\gamma$ in the screened Poisson problems. Our weighting strategy can improve the precision through the bias-variance trade-off.}
    \label{fig:ablation}
\end{figure*}

In this part, we study the effect of different weighting strategies and implementation details in section \ref{sec:detail}. \changed{Firstly, we present results in a simple 2D domain. As shown in Figure \ref{fig:2d_results} (a), for the simplest Laplace equation, the specially designed Poisson bound weight \cite{czekanski2024walking} presented better results than uniform weight and the proposed method. However, for the Poisson equation, the Poisson kernel does not provide a valid variance bound and hence underperforms compared to the proposed method and even uniform weight. We also present the MSE curves (Figure \ref{fig:ablation}) of 3D equations with the boundary defined by model 1 in Figure \ref{fig:models}.} For the 3D Poisson equation, the proposed method has a slight MSE improvement, but \changed{Poisson-bound weighting} increased the error since it cannot effectively estimate the variance of estimators. For the screened Poisson equation, the proposed weighting strategy filtered those highly-biased estimators, successfully reducing the bias brought by the approximated Green's function. 

\begin{figure*}
    \centering
    \includegraphics[width=0.95\textwidth]{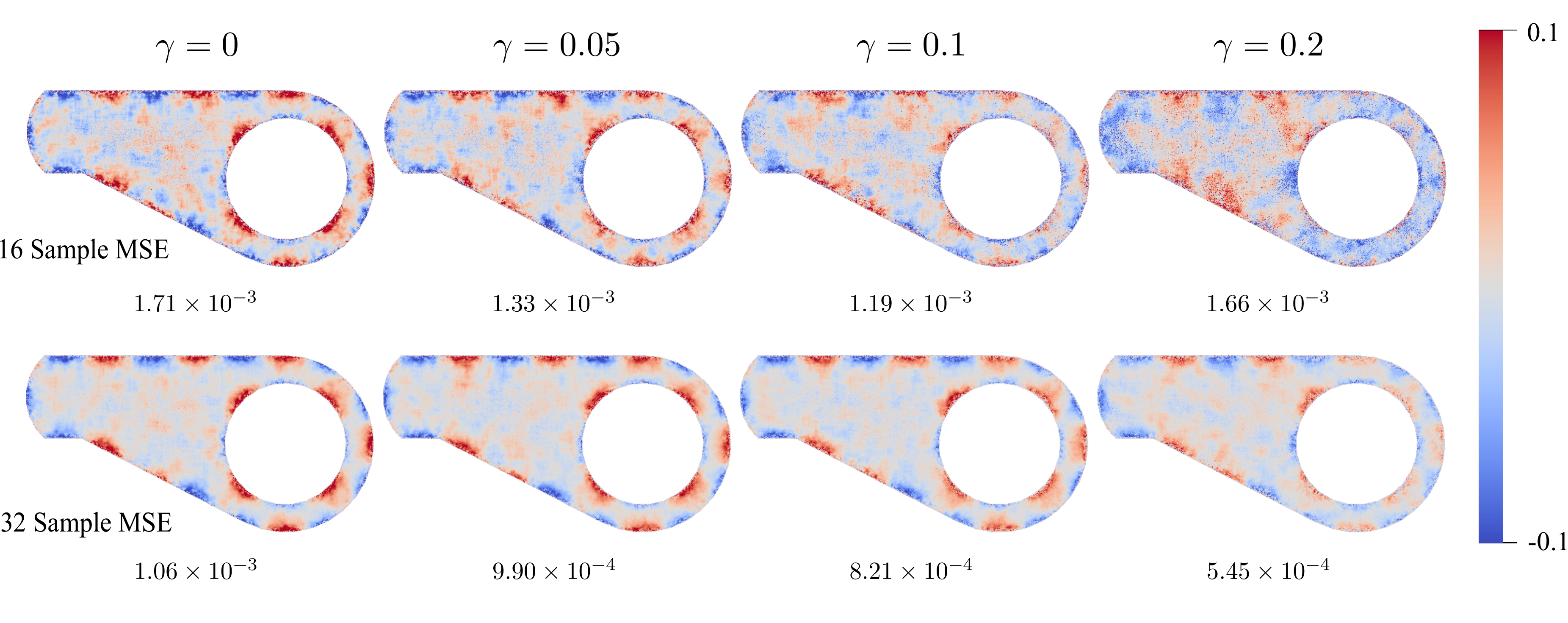}
    \caption{\textbf{Error map} under different value of $\gamma$. As $\gamma$ increases. We solve a screened Poisson problem with the proposed method with different $\gamma$. Larger $\gamma$ brings more noise, but reject those off-centered estimators with large bias.}
    \label{fig:graphical_abstract}
\end{figure*}

To examine the variance-bias trade-off, we visualize both numerical and visual results for the screened Poisson equation under varying values of the weighting parameter $\gamma$. As shown in Figure~\ref{fig:graphical_abstract}, when $\gamma = 0$ (corresponding to uniform weighting), the approximation of the Green's function introduces noticeable bias near the domain boundary. As $\gamma$ increases, bias is significantly reduced at the cost of more noise. After an increasing number of samples, the estimator with a larger $\gamma$ reached better results under the mean squared error metric, which is consistent with our analysis in section \ref{sec:weighting}.

We also present ablation results for our neighbor selection and importance strategy in Figure \ref{fig:ablation} (b). We compare the full algorithm with the naive neighbor selection ($\alpha = 0.99$) and the naive importance sampling.

\begin{figure*}
        \centering
    \includegraphics[width=0.85\textwidth]{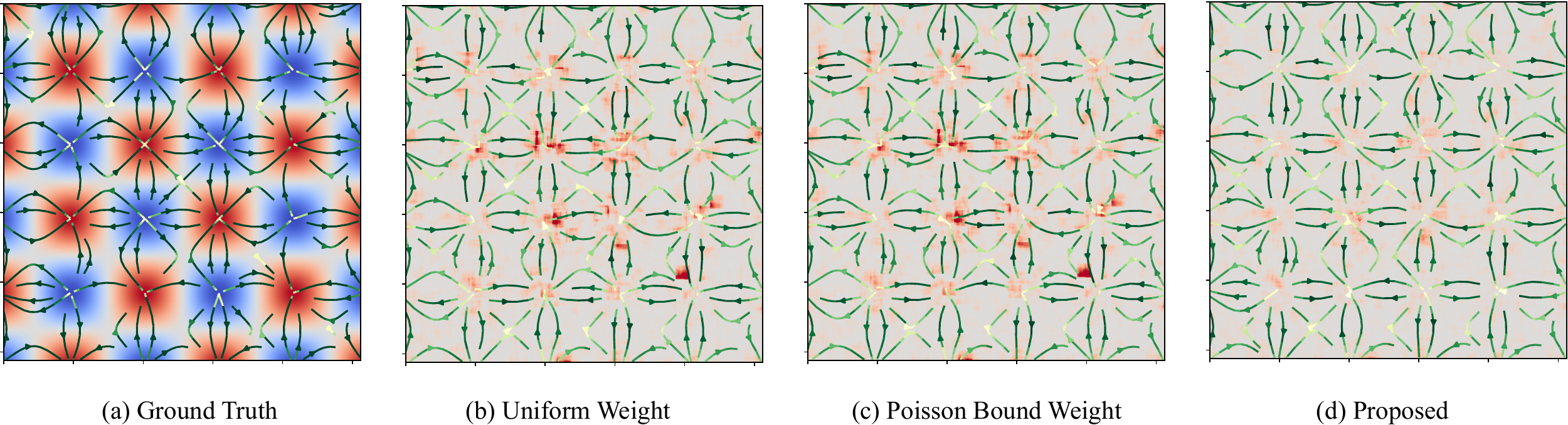}
    \caption{Illustration of the proposed method on the gradient estimation problem. (a) Ground truth \textbf{solution} and corresponding gradient field. (b) Estimating the solution gradient with uniform weight. (c) Estimating the solution gradient with Poisson bound weight. (d) Estimating the solution gradient with the proposed method. Background in (b), (c), and (d) is the \textbf{error map} of the gradient field. We can see significant correlation error in (b) and (c), which are eliminated by our filtering strategy.}
    \label{fig:gradient}
\end{figure*}

Compared with the solution estimation, gradient estimation commonly has a larger variance. Outlier samples will lead to remarkable artifacts. \changed{As shown in Figure \ref{fig:gradient}, the proposed weighting strategy successfully rejected the outlier sample and eliminated the correlation artifacts led by the outliers. Both uniform weighting and Poisson bound weighting cannot filter those samples with large errors since their weights are deterministic.}

\subsection{Compare with other Variance Reduction Methods}

\subsubsection{Dirichlet Problems}

We begin by evaluating our method and baselines on Dirichlet boundary value problems with the 3D Poisson equation.  Table~\ref{tab:poisson1} reports the mean squared error for each method under equal time budgets. In this case, our method consistently outperforms previous variance reduction methods. \changed{For all runs, we choose $\gamma = 0.05$ in our weighting strategy and keep $500$ seconds running time.}
  
\begin{table}[h]
    \centering
    \caption{Mean squared error of different variance reduction methods. Each major row corresponds to the models shown in Figure~\ref{fig:models}. Our method significantly improved the numerical precision under the MSE metric. In the high-frequency case, our MSE value is one magnitude smaller than the mean value caching. }
    \begin{tabular}{ccccc}
        \toprule
        $\omega$ & Vanilla & BVC & MVC & Proposed \\
        \midrule
        $\pi$ & $1.947 \times 10^{-4}$ & $1.848 \times 10^{-4}$ & $2.788 \times 10^{-5}$ & $\mathbf{4.504 \times 10^{-6}}$ \\
        $2\pi$ & $1.038 \times 10^{-3}$ & $5.949 \times 10^{-4}$ & $1.082 \times 10^{-4}$ & $\mathbf{1.302 \times 10^{-5}}$ \\
        $4\pi$ & $5.283 \times 10^{-3}$ & $9.547 \times 10^{-4}$ & $3.795 \times 10^{-4}$ & $\mathbf{4.352 \times 10^{-5}}$ \\
        \midrule
        $\pi$ & $6.915 \times 10^{-6}$ & $6.526 \times 10^{-5}$ & $3.437 \times 10^{-6}$ & $\mathbf{1.337 \times 10^{-6}}$ \\
        $2\pi$ & $2.886 \times 10^{-5}$ & $1.003 \times 10^{-4}$ & $1.432 \times 10^{-5}$ & $\mathbf{2.613 \times 10^{-6}}$ \\
        $4\pi$ & $1.159 \times 10^{-4}$ & $2.496 \times 10^{-4}$ & $4.002 \times 10^{-5}$ & $\mathbf{5.485 \times 10^{-6}}$ \\
        \midrule
        $\pi$ & $2.708 \times 10^{-5}$ & $3.106 \times 10^{-5}$ & $6.388 \times 10^{-6}$ & $\mathbf{8.707 \times 10^{-7}}$ \\
        $2\pi$ & $1.215 \times 10^{-4}$ & $6.256 \times 10^{-5}$ & $2.879 \times 10^{-5}$ & $\mathbf{2.806 \times 10^{-6}}$ \\
        $4\pi$ & $7.171 \times 10^{-4}$ & $1.707 \times 10^{-3}$ & $1.140 \times 10^{-4}$ & $\mathbf{9.209 \times 10^{-6}}$ \\
        \bottomrule
    \end{tabular}
    \label{tab:poisson1}
\end{table}

Following the same setup, we evaluate the screened Poisson equation using boundary and source terms derived from equation \ref{equ:poisson1}. As \citet{bakbouk2023mean} does not address screened Poisson equations, we only compare against the boundary value caching method \cite{miller2023boundary}. We fix $\gamma = 0.3$ for all evaluations. Table~\ref{tab:screened_poisson} shows that our method consistently outperforms boundary value caching across various frequencies $\omega$.

\begin{table}[h]
    \centering
    \caption{Numerical results on screened Poisson problems. Our method gets a 3 to 5 times improvement under the MSE metric.}
    \begin{tabular}{ccccc}
        \toprule
        $\omega$ & $\sigma$ & BVC & Vanilla & Proposed \\
        \midrule
        $\pi$  & $5$ & $9.945 \times 10^{-4}$ & $9.844 \times 10^{-4}$ & $\mathbf{8.528 \times 10^{-5}}$ \\
        $2\pi$ & $5$ & $1.205 \times 10^{-3}$ & $3.430 \times 10^{-3}$ & $\mathbf{2.616 \times 10^{-4}}$ \\
        $4\pi$ & $5$ & $1.016 \times 10^{-2}$ & $2.028 \times 10^{-2}$ & $\mathbf{5.034 \times 10^{-4}}$ \\
        \midrule
        $\pi$  & $5$ & $1.393 \times 10^{-4}$ & $3.711 \times 10^{-5}$ & $\mathbf{1.713 \times 10^{-5}}$ \\
        $2\pi$ & $5$ & $2.206 \times 10^{-4}$ & $1.265 \times 10^{-4}$ & $\mathbf{3.586 \times 10^{-5}}$ \\
        $4\pi$ & $5$ & $5.107 \times 10^{-4}$ & $5.437 \times 10^{-4}$ & $\mathbf{7.880 \times 10^{-5}}$ \\
        \midrule
        $\pi$  & $5$ & $4.201 \times 10^{-5}$ & $8.373 \times 10^{-5}$ & $\mathbf{2.688 \times 10^{-5}}$ \\
        $2\pi$ & $5$ & $6.394 \times 10^{-5}$ & $3.086 \times 10^{-4}$ & $\mathbf{5.624 \times 10^{-5}}$ \\
        $4\pi$ & $5$ & $8.170 \times 10^{-4}$ & $2.135 \times 10^{-3}$ & $\mathbf{1.481 \times 10^{-4}}$ \\
        \bottomrule
    \end{tabular}
    \label{tab:screened_poisson}
\end{table}

\subsubsection{Mixed Boundary Condition}

Our method also generalizes to problems involving mixed boundary conditions. In this experiment, we validate its performance on problems with both Dirichlet and Neumann boundaries. Results are presented in Table \ref{tab:screened_poisson_mixed}. Our method reached the best result on boundary data 1 and 2. In the model with the largest Neumann boundary, boundary value caching performs better. 

\begin{table}[h]
    \centering
    \caption{Mean squared error of different variance reduction methods on a screened Poisson problem with mixed boundary conditions. The proposed approach has better MSE except the Model 3, which has $50\%$ Neumann boundary.}

    \begin{tabular}{ccccc}
        \toprule
        $\omega$ & $\sigma$ & BVC & Vanilla & Proposed \\
        \midrule
        $\pi$  & $5$ & $1.401 \times 10^{-3}$ & $3.356 \times 10^{-3}$ & $\mathbf{3.886 \times 10^{-4}}$ \\
        $2\pi$ & $5$ & $1.232 \times 10^{-3}$ & $1.133 \times 10^{-2}$ & $\mathbf{6.528 \times 10^{-4}}$ \\
        $4\pi$ & $5$ & $1.034 \times 10^{-2}$ & $7.717 \times 10^{-2}$ & $\mathbf{2.175 \times 10^{-3}}$ \\
        \midrule
        $\pi$  & $5$ & $1.499 \times 10^{-4}$ & $2.406 \times 10^{-4}$ & $\mathbf{5.116 \times 10^{-5}}$ \\
        $2\pi$ & $5$ & $3.803 \times 10^{-4}$ & $8.974 \times 10^{-4}$ & $\mathbf{1.346 \times 10^{-4}}$ \\
        $4\pi$ & $5$ & $6.774 \times 10^{-4}$ & $3.607 \times 10^{-3}$ & $\mathbf{4.668 \times 10^{-4}}$ \\
        \midrule
        $\pi$  & $5$ & $\mathbf{2.907 \times 10^{-4}}$ & $1.193 \times 10^{-2}$ & $1.595 \times 10^{-3}$ \\
        $2\pi$ & $5$ & $\mathbf{1.372 \times 10^{-4}}$ & $3.042 \times 10^{-2}$ & $1.741 \times 10^{-4}$ \\
        $4\pi$ & $5$ & $\mathbf{3.518 \times 10^{-3}}$ & $2.862 \times 10^{-1}$ & $1.928 \times 10^{-2}$ \\
        \bottomrule
    \end{tabular}
    \label{tab:screened_poisson_mixed}
\end{table}

\subsubsection{Convergence Curves}

\begin{figure*}
    \centering
    \subfigbottomskip=2pt
    \subfigcapskip=-5pt
        \subfigure[Poisson, Dirichlet, $\omega = \pi$]{
        \includegraphics[width=0.3\linewidth]{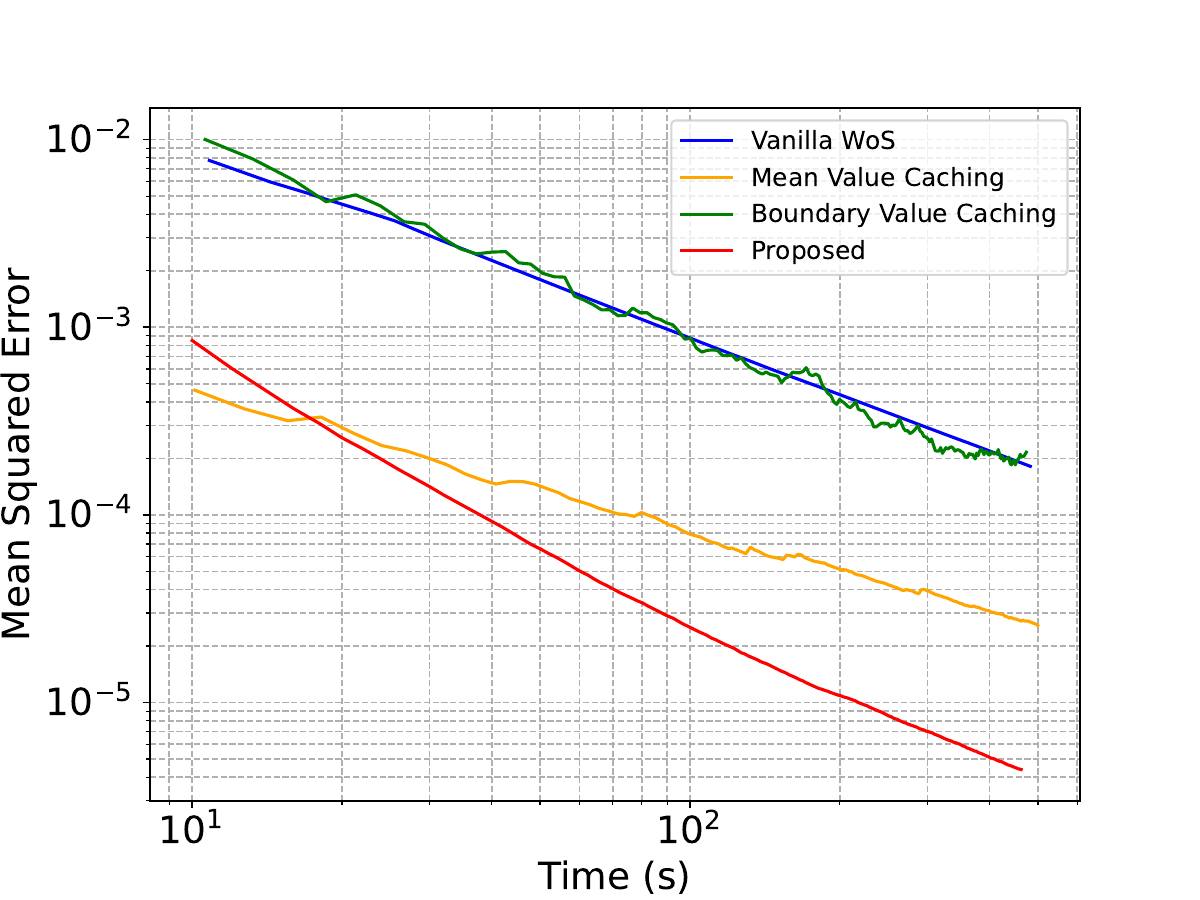}}
    \subfigure[Poisson, Dirichlet, $\omega = 2\pi$]{
        \includegraphics[width=0.3\linewidth]{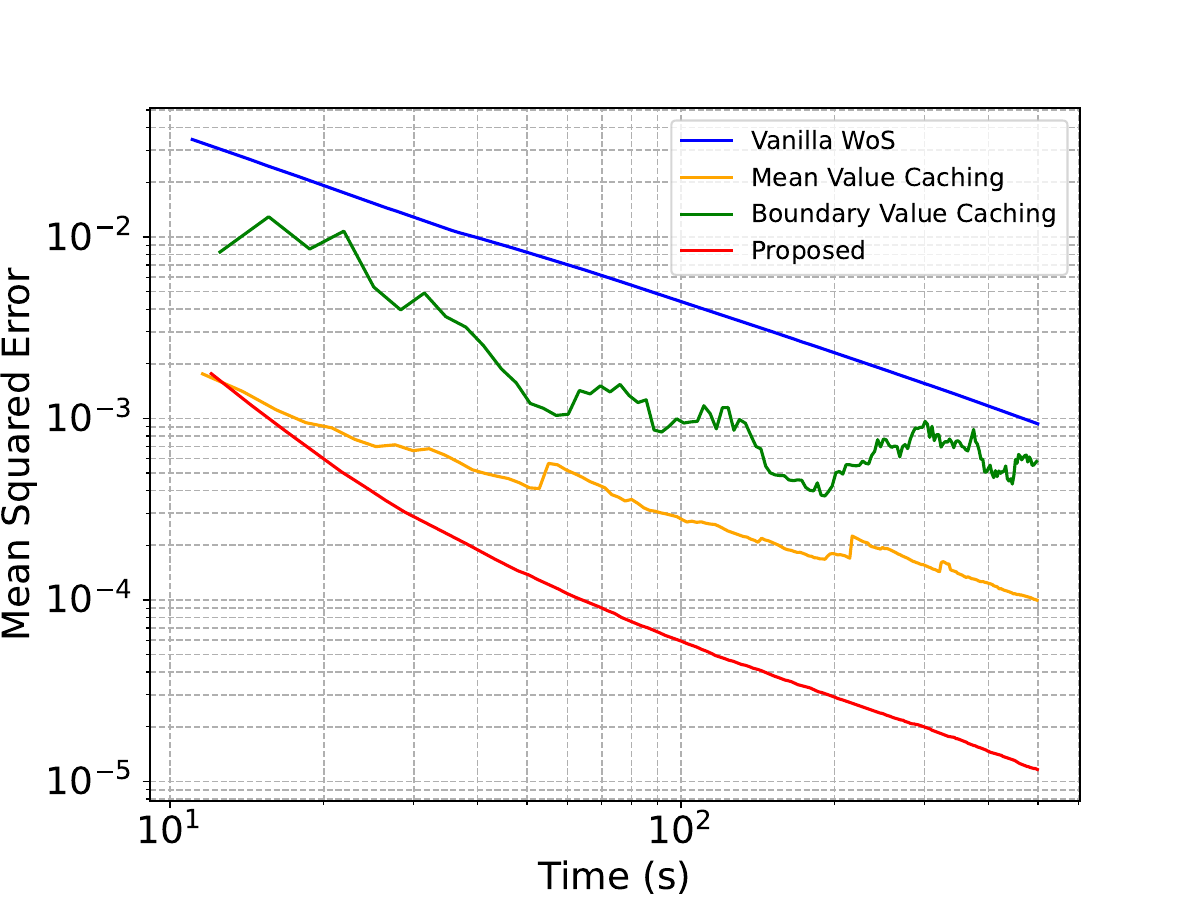}} 
    \subfigure[Poisson, Dirichlet, $\omega = 4\pi$]{
        \includegraphics[width=0.3\linewidth]{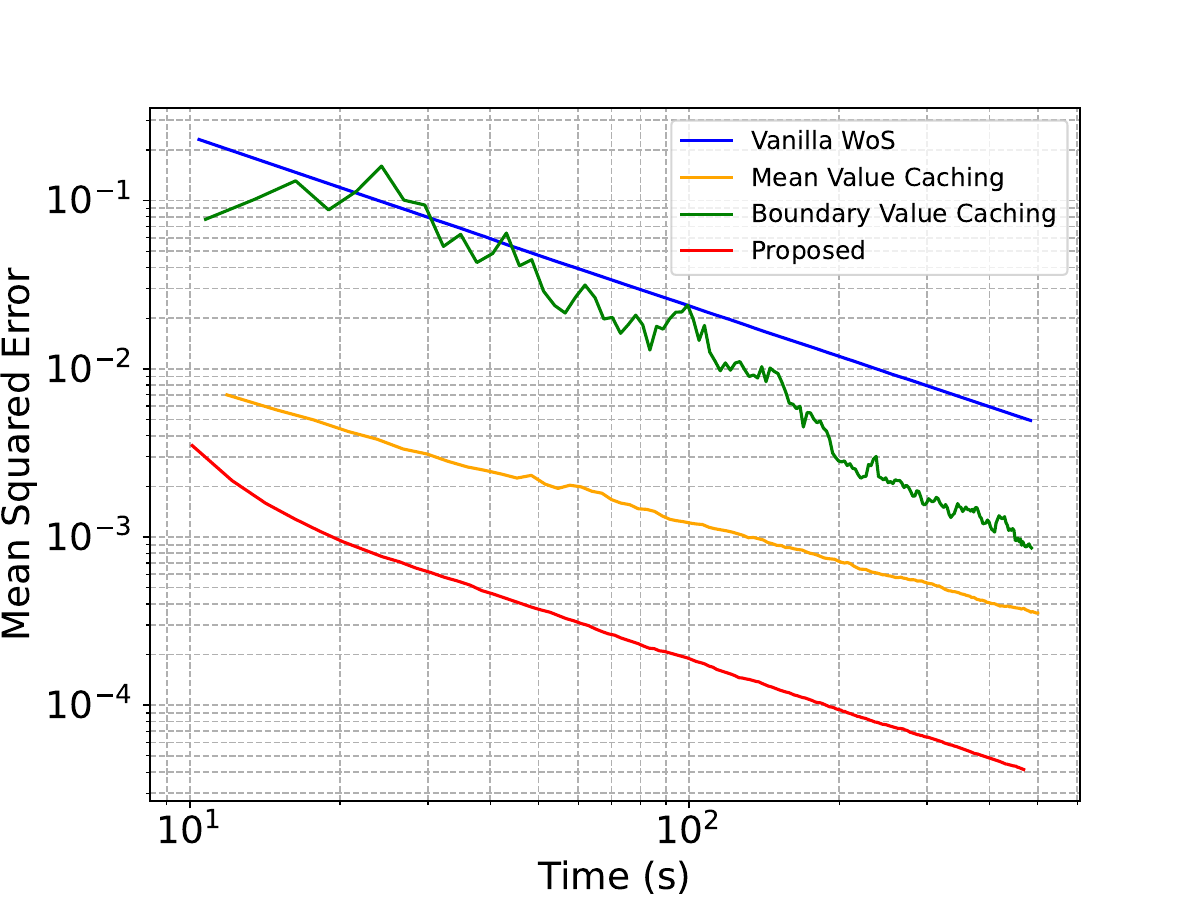}} \\
    \subfigure[Poisson, Mixed, $\omega = \pi$]{
        \includegraphics[width=0.3\linewidth]{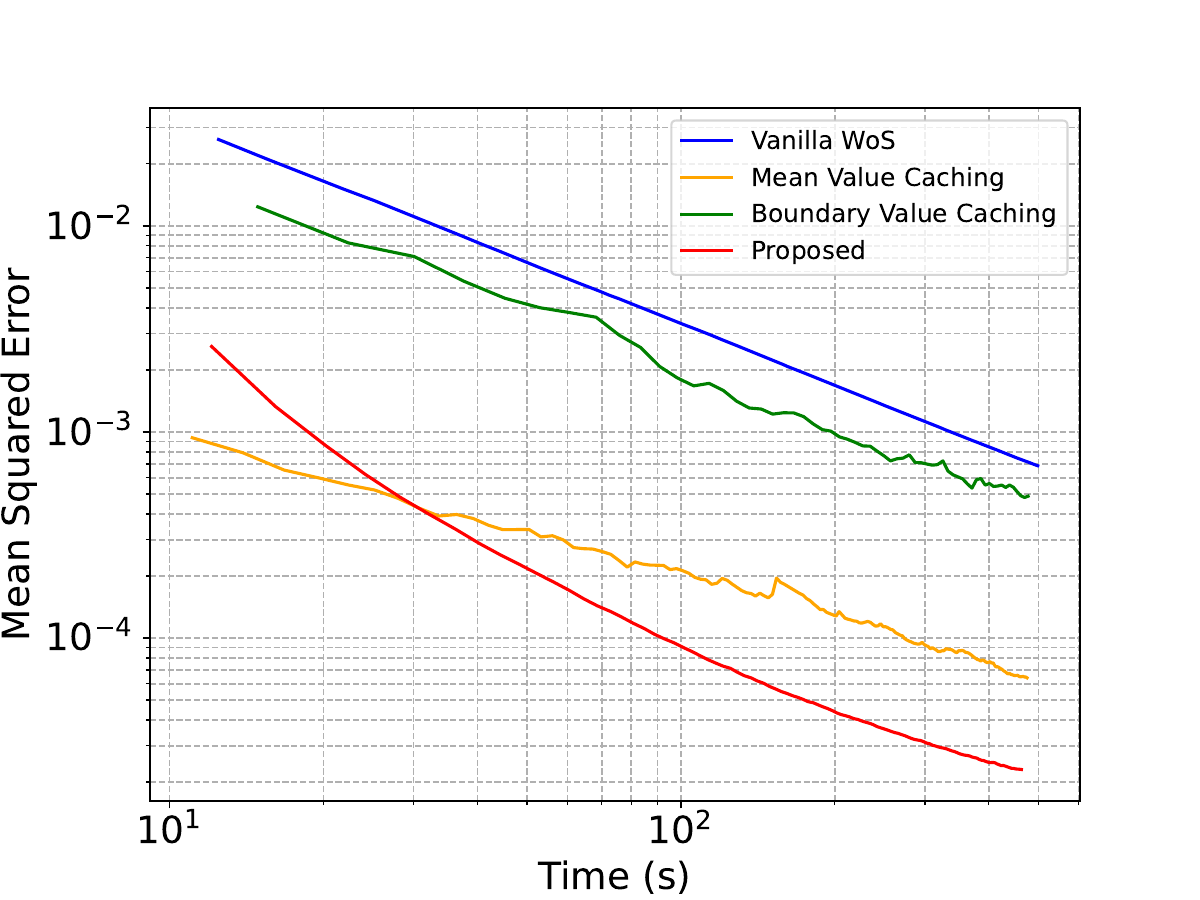}} 
    \subfigure[Poisson, Mixed, $\omega = 2\pi$]{
        \includegraphics[width=0.3\linewidth]{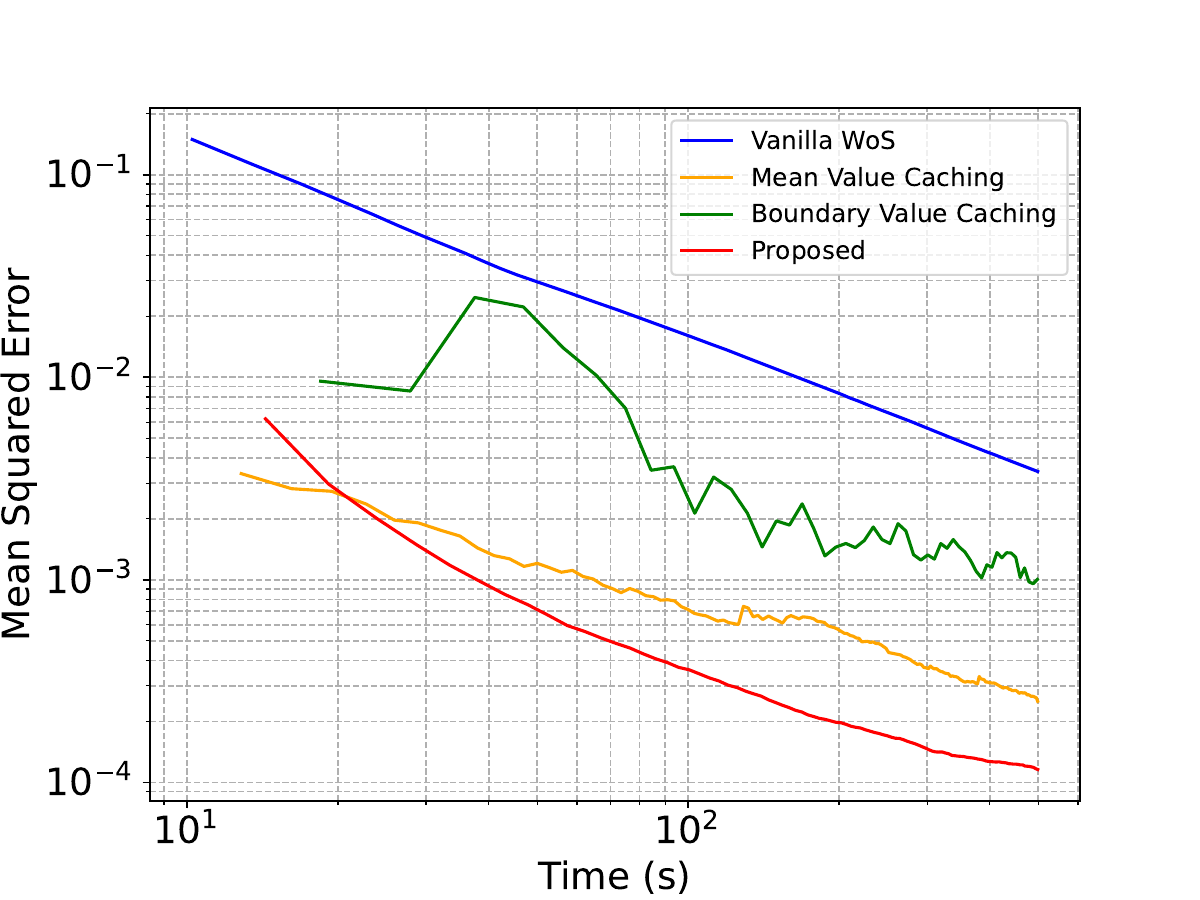}}
    \subfigure[Poisson, Mixed, $\omega = 4\pi$]{
        \includegraphics[width=0.3\linewidth]{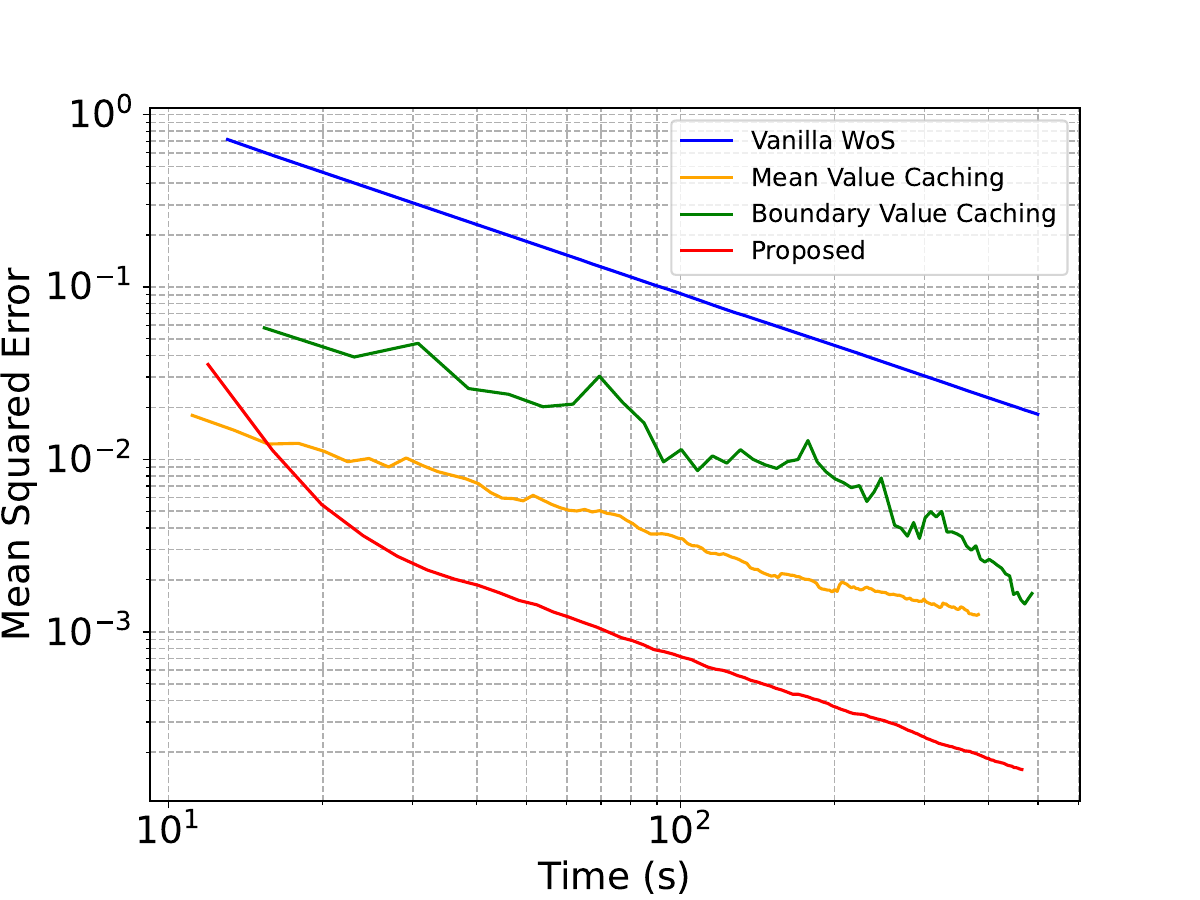}} \\
    \subfigure[Screened Poisson, Dirichlet, $\omega = \pi$]{
        \includegraphics[width=0.3\linewidth]{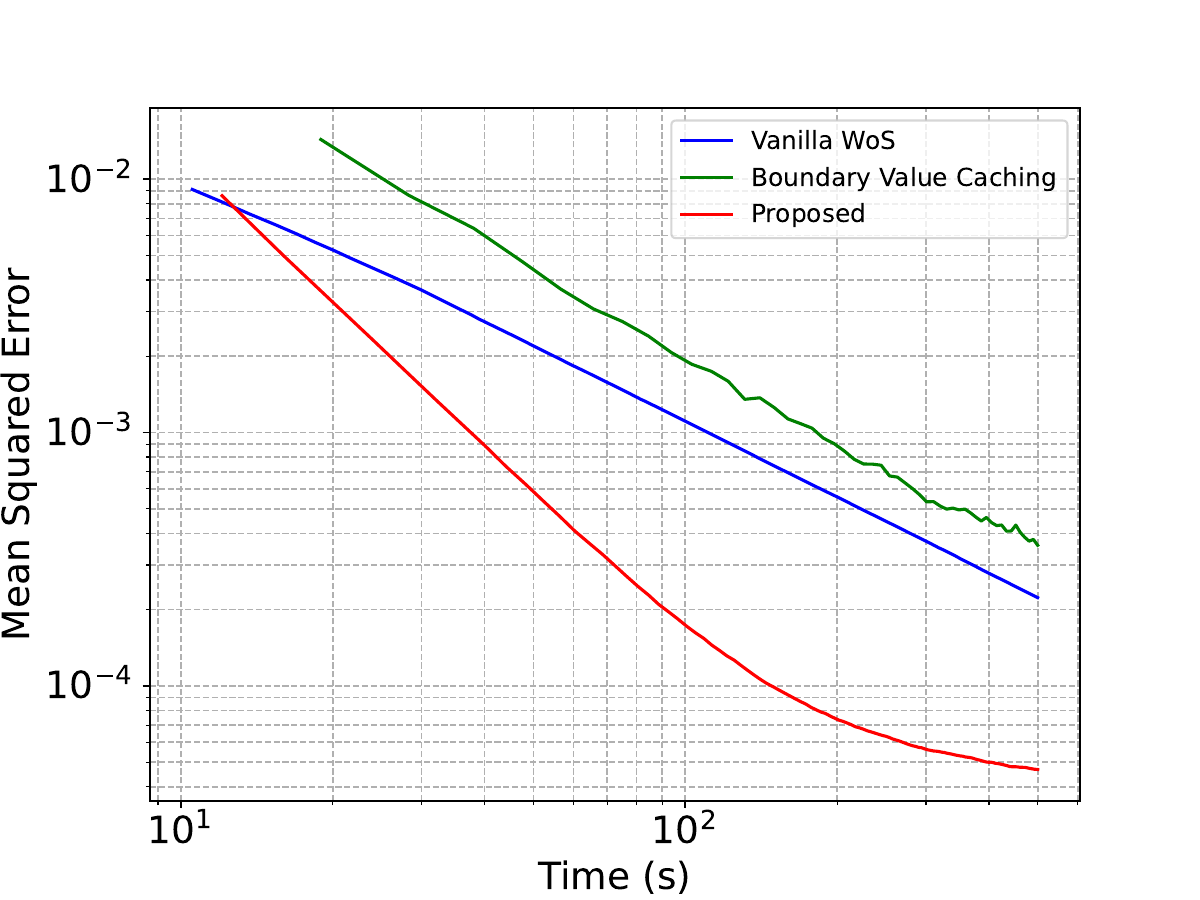}}
    \subfigure[Screened Poisson, Dirichlet, $\omega = 2\pi$]{
        \includegraphics[width=0.3\linewidth]{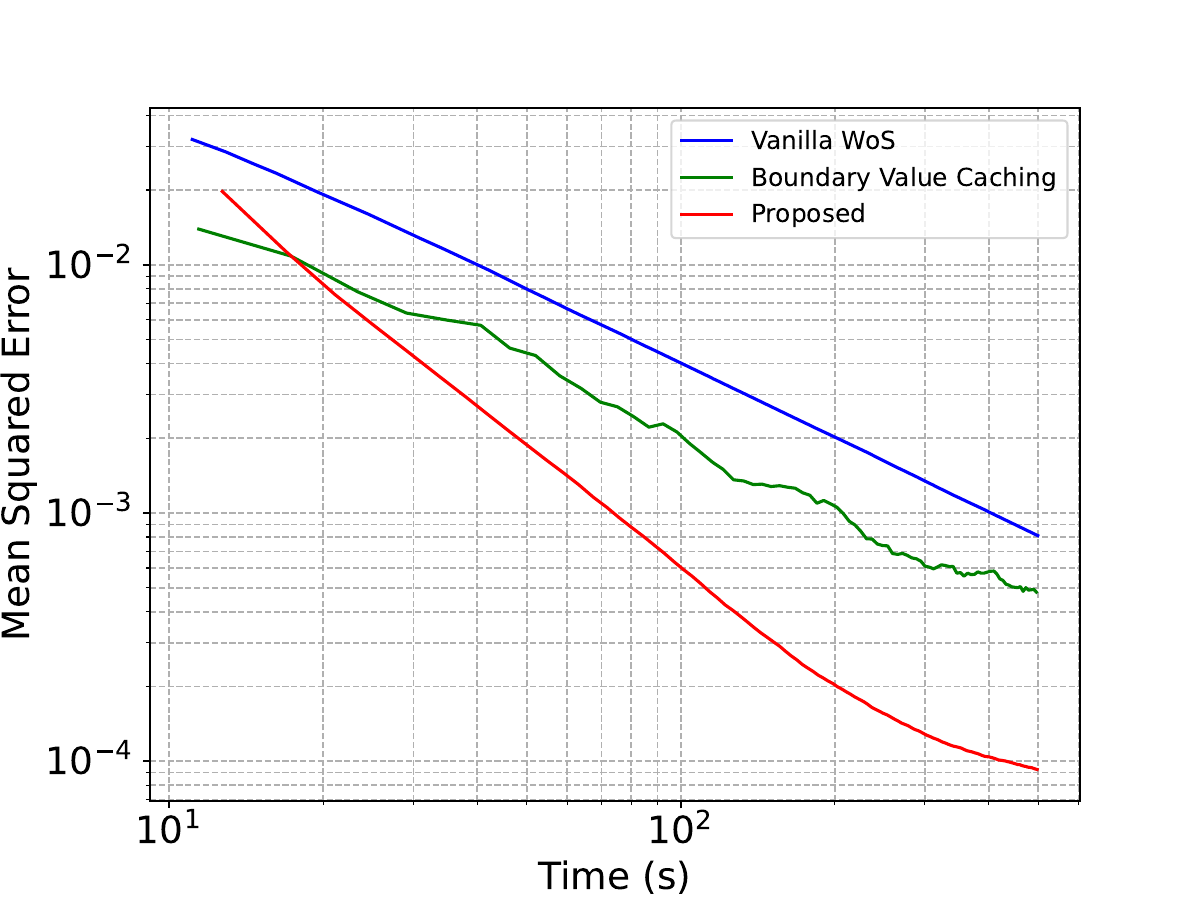}} 
    \subfigure[Screened Poisson, Dirichlet, $\omega = 4\pi$]{
        \includegraphics[width=0.3\linewidth]{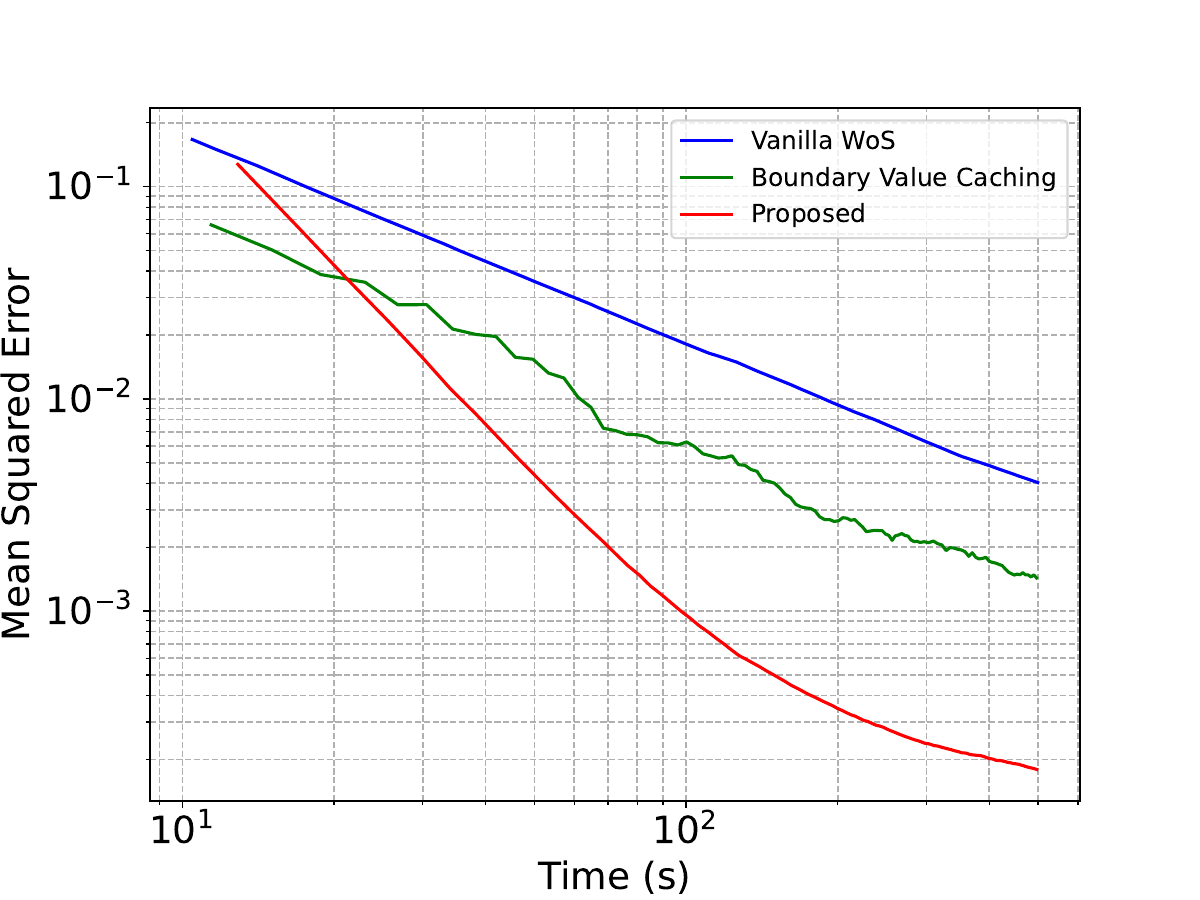}} \\
    \subfigure[Screened Poisson, Mixed, $\omega = \pi$]{
        \includegraphics[width=0.3\linewidth]{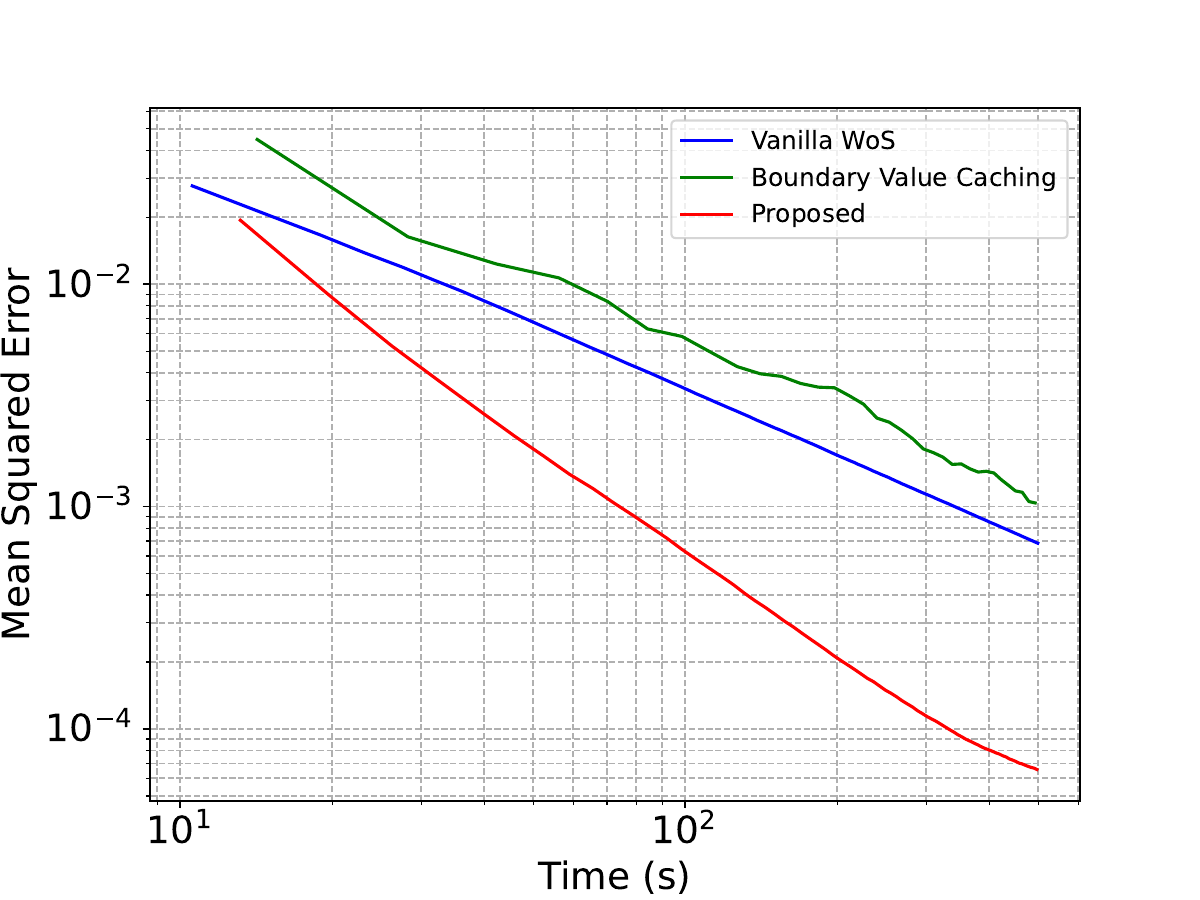}} 
    \subfigure[Screened Poisson, Mixed, $\omega = 2\pi$]{
        \includegraphics[width=0.3\linewidth]{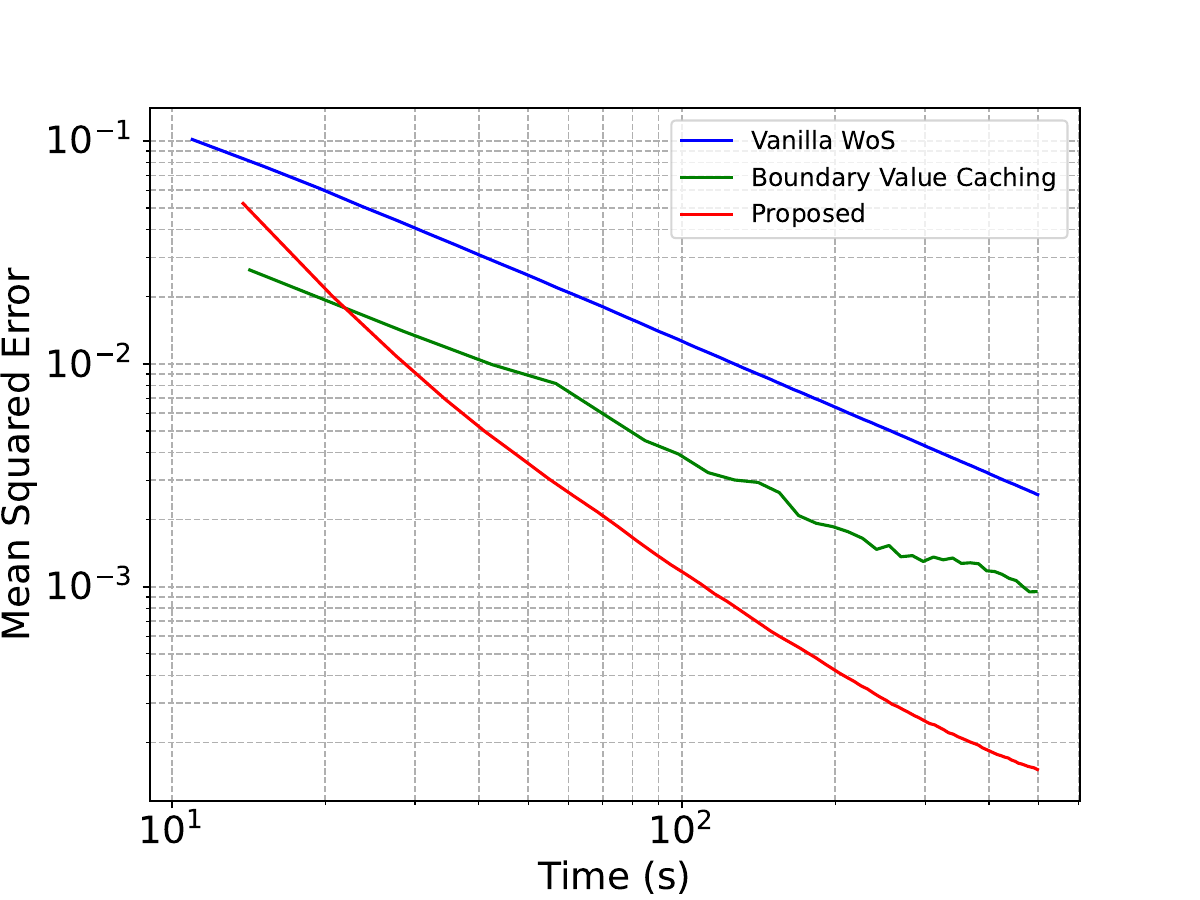}}
    \subfigure[Screened Poisson, Mixed, $\omega = 4\pi$]{
        \includegraphics[width=0.3\linewidth]{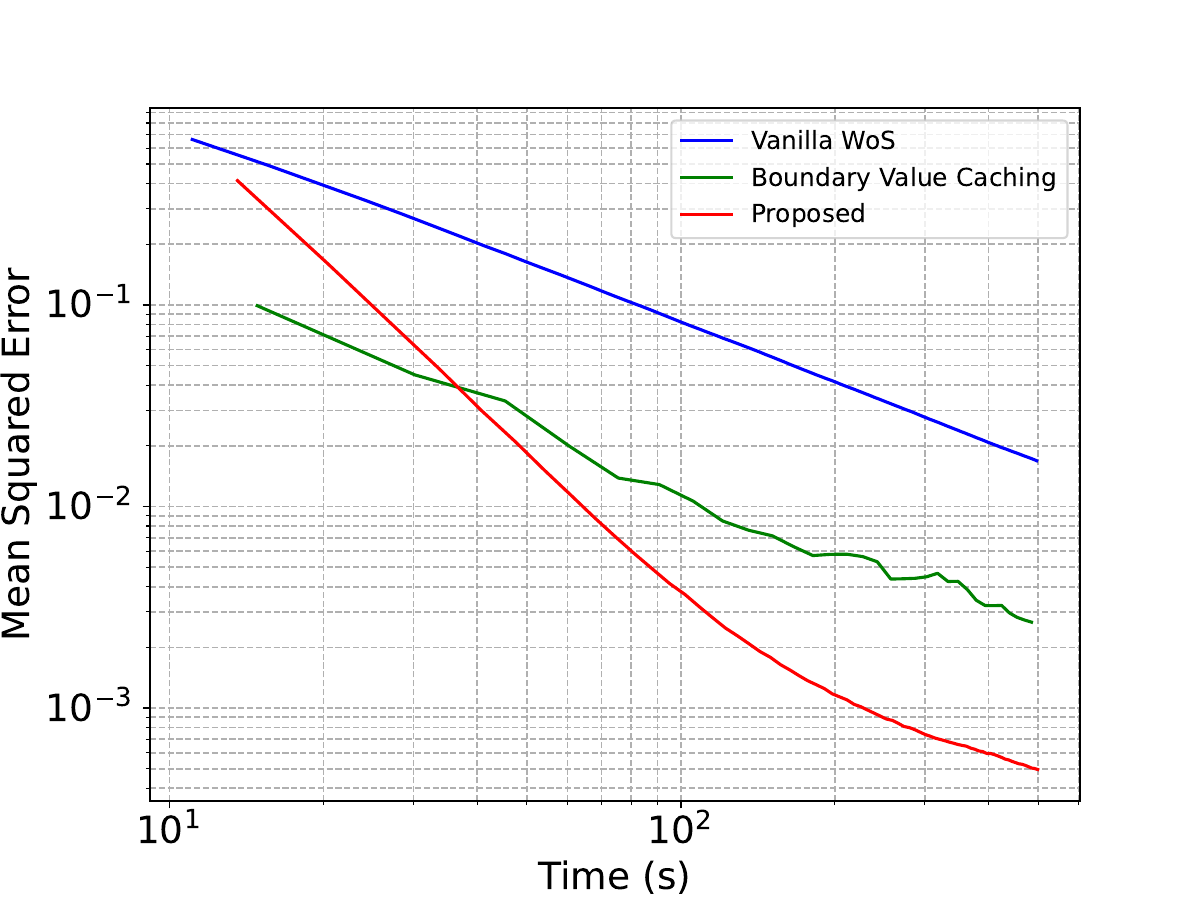}} \\
    \caption{Convergence curves comparing the proposed method with existing variance reduction techniques. The plot shows the mean squared error (MSE) over time for vanilla Walk on Spheres, mean value caching, boundary value caching, and the proposed statistically-weighted off-centered estimator. Our method achieves consistently lower MSE across sampling budgets, demonstrating faster convergence and improved accuracy, particularly in high-frequency or biased settings.}
    \label{fig:time_mse}
\end{figure*}

We present convergence curves of different algorithms under different equation types, boundary conditions, and ground truth frequencies in Figure \ref{fig:time_mse}. \changed{All runs are in 500 seconds, and the geometry is model 1 in figure \ref{fig:models}.}

% \subsubsection{Gradient Estimation}

% Finally we report results of ablation study

\section{Limitations and Future Work}

\textit{Limitations.} \changed{We introduced a new statistical filtering method to WoS-type solvers, solving the correlation artifact and bias issues in combining off-centered estimators. While handling these problems, our method has some known limitations. As stated in section \ref{sec:weighting}, our method is no longer unbiased. In the early sampling stage, since the estimates are noisy, our results may generate worse results. It is also worth mentioning that, compared with standard solvers, the off-centered scheme does not keep the mean value property. For simple Laplace equations, the final result may be larger/smaller than the boundary maxima/minima.}

\textit{Adaptive Strategies.} \changed{In our current implementation, the hyperparameters in the weighting strategy and the neighbor selection strategy are all fixed. The number of walks from each evaluation point is equal. Making these hyperparameters adaptive is an important future direction.}

\textit{Denoising}. \changed{Similar to the development of Monte Carlo rendering, denoising techniques can also be used with stochastic PDE solvers. Although we can directly use the image denoisers used in Monte Carlo rendering, a denoising method specialized for the PDE solvers has not been thoroughly explored.}

\textit{Pointwise estimators.}  
\changed{Our method is compatible with various pointwise estimators. While traditional walk-on-spheres (WoS) algorithms~\cite{sawhney2020monte, sawhney2023walk} employ specific estimators at each step, alternative schemes such as the walk-on-boundary (WoB) algorithm~\cite{sugimoto2023practical} can also be integrated. Furthermore, enhancements like those proposed by~\citet{huang2025path} are orthogonal to our approach and can be seamlessly integrated within the same estimator reuse mechanism.}

\section{Conclusion}

We have presented a novel statistically-weighted off-centered estimator \changed{for WoS-type PDE solvers} in computer graphics. By integrating a similarity-based weighting scheme, our method reduces correlation artifacts and allows for controlled bias when approximate Green’s functions are used. This framework generalizes existing off-centered approaches and extends their applicability to a broader class of PDEs and boundary conditions, including the screened Poisson equation and gradient field estimation.

Our experiments show that the proposed method achieves significant variance reduction compared to both standard \changed{WoS-type} PDE solvers and prior correlation-based techniques. The resulting solver demonstrates robust convergence properties across a wide range of scenarios, offering an effective and general-purpose tool for solving PDEs in graphics applications.

\begin{acks}
The authors thank the anonymous reviewers for their valuable feedback. This research was supported by the National Key Research and Development Program of China (Grant No.2023YFB3309000) and Ministry of Industry and Information Technology of the People's Republic of China (Grant 20241670004, WTH2025104). 
\end{acks}

%
% The next two lines define the bibliography style to be used, and
% the bibliography file.
\bibliographystyle{ACM-Reference-Format}
% \bibliography{sample-base}
\bibliography{ref}

% \newpage

\end{document}